\documentclass{article}

\usepackage{arxiv}
\usepackage{algorithm}
\usepackage{algpseudocode}
\usepackage[utf8]{inputenc} 
\usepackage[T1]{fontenc}    
\usepackage{hyperref}       
\usepackage{url}            
\usepackage{booktabs}       
\usepackage{amsfonts}       
\usepackage{nicefrac}       
\usepackage{microtype}      
\usepackage{graphicx}
\usepackage{multicol}
\usepackage{float}
\usepackage{newunicodechar}
\newunicodechar{−}{\textminus}
\newunicodechar{≤}{\leq}
\usepackage{doi}
\usepackage{subcaption}
\usepackage{tabularx}
\usepackage{longtable}
\usepackage{siunitx}

\usepackage{caption}    
\usepackage{bm}  
\usepackage{amsmath}
\usepackage[nameinlink, noabbrev]{cleveref} 
\usepackage{natbib}
\usepackage{scalerel}

\usepackage{multirow}
\usepackage{enumitem}   

\title{GraphRAG for Engineering Diagrams: ChatP\&ID Enables LLM Interaction with P\&IDs}

\author{ 
    \href{https://orcid.org/0000-0001-6382-8196}{\includegraphics[scale=0.06]{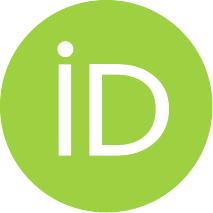}\hspace{1mm}Achmad Anggawirya Alimin} \\
	Process Intelligence Research Group\\
	Department of Chemical Engineering\\
	Delft University of Technology\\
	\And
    \href{https://orcid.org/0000-0001-8885-6847}{\includegraphics[scale=0.06]{figures/orcid.pdf}\hspace{1mm}Artur M. Schweidtmann}\thanks{corresponding author} \\
	Process Intelligence Research Group\\
	Department of Chemical Engineering\\
	Delft University of Technology\\
    \texttt{A.Schweidtmann@tudelft.nl} \\
}

\renewcommand{\shorttitle}{ChatP\&ID}

\hypersetup{
pdftitle={ChatP\&ID},
pdfauthor={A. Anggawirya Alimin},
}

\begin{document}
\maketitle

\begin{abstract}
Large Language Models (LLMs) combined with Retrieval-Augmented Generation (RAG) and knowledge graphs offer new opportunities for interacting with engineering diagrams such as Piping and Instrumentation Diagrams (P\&IDs). However, directly processing raw images or smart P\&ID files with LLMs is often costly, inefficient, and prone to hallucinations.
This work introduces ChatP\&ID, an agentic framework that enables grounded and cost-effective natural-language interaction with P\&IDs using Graph Retrieval-Augmented Generation (GraphRAG), a paradigm we refer to as GraphRAG for engineering diagrams.
Smart P\&IDs encoded in the DEXPI standard are transformed into structured knowledge graphs, which serve as the basis for graph-based retrieval and reasoning by LLM agents. This approach enables reliable querying of engineering diagrams while significantly reducing computational cost.
Benchmarking across commercial LLM APIs (OpenAI, Anthropic) demonstrates that graph-based representations improve accuracy by 18\% over raw image inputs and reduce token costs by 85\% compared to directly ingesting smart P\&ID files. While small open-source models still struggle to interpret knowledge graph formats and structured engineering data, integrating them with VectorRAG and PathRAG improves response accuracy by up to 40\%. Notably, GPT-5-mini combined with ContextRAG achieves 91\% accuracy at a cost of only \$0.004 per task. 
The resulting ChatP\&ID interface enables intuitive natural-language interaction with complex engineering diagrams and lays the groundwork for AI-assisted process engineering tasks such as Hazard and Operability Studies~(HAZOP) and multi-agent analysis.
\end{abstract}


\section{Introduction}
\label{sec:introduction}

Piping and Instrumentation Diagrams (P\&IDs) serve as an essential source of information in process engineering~\citep{toghraei2019piping}. Engineers rely on accurate interpretation of P\&ID data throughout the lifecycle of a plant, including design, operation, maintenance, expansion, and risk assessments. However, interacting with P\&IDs remains a bottleneck due to their complex structure and the heavy reliance on manual workflow. Current practices typically involve manually tracing process lines and equipment in PDF or Computer-Aided Engineering (CAE) files, a process that is both time-consuming and error-prone.

Recent digitalization efforts in process engineering are shifting P\&ID development toward a more data-centric paradigm. “Smart” or “intelligent” P\&IDs embed database structures within traditional diagrams, enabling easier data exchange and more automated workflows. While most industrial P\&IDs still exist as PDFs or physical drawings, ongoing research is converting legacy assets into smart, machine-readable formats~\citep{theisen2023digitization}. Standardization initiatives such as DEXPI have also been instrumental in this shift by defining a unified data schema for P\&IDs~\citep{Theissen2021}. Leveraging these standards, our group developed pyDEXPI, a Python implementation of the DEXPI information model~\citep{Goldstein2025}. Although these advances lay a strong foundation for the future of process data management, practical solutions that seamlessly integrate into engineers’ existing workflows remain unexplored. There is a clear opportunity to unlock more value from digital P\&IDs.

Lately, leveraging generative AI for P\&IDs has become a central focus for advancing AI in process engineering~\citep{schweidtmann2024generative}. Several studies have explored the use of AI to extract information from or analyze P\&IDs. A common approach feeds P\&ID images directly into Large Language Models~(LLMs) and asks for tasks ranging from component identification~\citep{medhane2025automated} to HAZOP analysis~\citep{LEE2026107039, mukharror2025use}. However, this approach obscures the image recognition steps, limiting transparency into how the LLM extracts information. This will make systematic improvement difficult, other than changing the LLM model. Another method passes the smart P\&ID files as textual context for the LLM to read and modify~\citep {gowaikar2024agentic}. While more transparent than images, this approach remains prone to hallucinations and is challenging for engineers to inspect, as the XML is designed for CAD (Computer-Aided Design) software, which lacks semantic readability. They can also be resource-intensive; in one study~\citep{alimin2025talking}, a single simple P\&ID page required over 150,000 tokens. Despite this exploration, to date, the optimal representation of P\&IDs for AI-based flowsheet analysis remains an open question.

In our previous work, we established a foundation for representing P\&IDs as knowledge graphs and enabling LLMs to interpret these engineering diagrams~\citep{alimin2025talking}. This knowledge graph representation offers a more intuitive and interactive way for process engineers to engage with P\&ID data by passing graph context to LLMs. The latest advances in graph reasoning methods have also revealed substantial untapped opportunities to leverage graph-based representations for more powerful and accurate querying. For example, recent work has shown that LLMs can extract localized information and summarize global graph semantics~\citep{edge2024local}, and several studies explore iterative graph exploration guided by LLM reasoning~\citep{sun2023think, ma2024think, he2024give}. Furthermore, advances in LLM-generated Cypher queries showcase the potential for natural-language-driven retrieval from Labeled Property Graphs~(LPG)~\citep{szlobodnyik5669662feedback, gusarov2025multi, gupta2025pidqa}. Yet, none of these GraphRAG techniques have been evaluated on P\&IDs, and their performance has not been benchmarked in terms of computational efficiency, cost, and response accuracy when compared to image-based interpretation or direct ingestion of smart P\&IDs across different LLMs. To the best of our knowledge, this is the first application of GraphRAG to structured engineering diagrams such as Piping and Instrumentation Diagrams. 

To address these gaps, this work extends our previous methodology by integrating state-of-the-art graph-based query methods for P\&ID. We also provide a comprehensive evaluation of multiple online and offline LLMs and model scales, benchmarking them across response accuracy, computational time, and cost. Finally, we present our work as a chat interface called \textbf{ChatP\&ID}, an agent-driven chat interface in which LLMs autonomously select and invoke GraphRAG tools to query information from the P\&ID. This enables process engineers to interact with and extract insights from P\&IDs in an intuitive, efficient way, while ensuring the accuracy, transparency, and traceability of the retrieved information. More broadly, we aim to demonstrate how advances in knowledge graph technology can drive practical impact in process engineering. In the long term, the integration of P\&IDs into foundation models represents a foundational capability for future GenAI applications, including automated P\&ID correction, AI-assisted HAZOP studies~\citep{schweidtmann2024generative}, and Multi-Agent System~(MAS) for process engineering workflow~\citep{rupprechtMultiagentSystemsChemical2025}.

\section{Preliminaries}
\label{sec:preliminaries}
This section introduces the foundational concepts and technologies that enable our proposed framework. 

\subsection{Representing P\&ID as knowledge graphs}
\label{subsec: Representing P&ID as knowledge graphs}
P\&IDs serve as the detailed blueprint for a chemical process facility, describing equipment, piping, control logic, and safety elements in accordance with standardized industrial conventions \citep{toghraei2019piping}. Despite their central role in engineering, safety, and operational workflows, P\&ID data is often stored in static formats such as images or PDFs, making updates and information retrieval labor-intensive and error-prone. To address this limitation, the DEXPI initiative defines a standardized semantic data model that enables consistent digital representation and interoperability across CAD tools~\citep{Theissen2021}. A P\&ID inherently contains two categories of information: (i) topological connectivity describing how entities interact (e.g., a pump feeding a tank, or a level indicator measuring tank level), and (ii) specification data describing attributes such as equipment materials, ratings, etc. While DEXPI captures these elements, its default representation is designed to be read by the CAD program, which lacks semantic expressiveness for Machine Learning~(ML) and generative AI implementation. Therefore, we implement pyDEXPI, a Python library that enhances DEXPI's usability by enabling the extraction and export of P\&ID data to formats such as tables, JSON, and, most importantly, knowledge graphs~\citep{Goldstein2025}. 

A graph can be represented as an index-free adjacency database consisting of nodes and edges \citep{latora2017complex}. Nodes represent entities in the dataset, while edges define relationships between them. Due to their flexible structure and intuitive query language, graph databases are widely used in applications such as social networking, recommender systems, and fraud detection. Knowledge graphs also demonstrated their effectiveness in process engineering for modeling entire systems, such as P\&IDs~\citep{morbach2007ontocape,eibeck_j-park_2019}. This approach represents equipment and instrumentation as nodes and their relationships as semantically meaningful edges. This not only provides a richer and more intuitive abstraction of P\&ID content that closely aligns with how engineers conceptualize process interactions, but also enables machine-readability, modification, and integration with language models \citep{alimin2025talking}.

There are two graph data formats commonly used for knowledge graph representation: (i) the Resource Description Framework~(RDF), which represents information as a triple store consisting of (subject, predicate, object) statements~\citep{pan2009resource}. In the context of a P\&ID, this could be expressed as \texttt{(Pump1, hasPressure, 50psi)}. RDF is a W3C standard and is typically queried using the SPARQL query language. (ii) The Labeled Property Graph~(LPG) model represents information using nodes, edges, and associated properties that can be assigned to both nodes and relationships~\citep{barrasa2023building}. In a P\&ID, this could be represented as a node \texttt{(id: 1, tag: "Pump1", type: "CentrifugalPump", pressure: 50, pressure\_unit: "psi")} connected to a valve node via a labeled edge: \texttt{(Pump1)-[: CONNECTED\_TO {flow: 10}]->(Valve2)}. LPG-based graphs are typically queried using languages such as Cypher or Gremlin and are commonly implemented in graph databases such as Neo4j. While each of these formats has its own advantages~\citep{baken2020linked, di2023lpg}, the choice of representation for a P\&ID should be guided by the specific requirements of the implementation context.

\subsection{Graph-based Retrieval Augmented Generation~(GraphRAG)}
\label{subsec: Graph-based Retrieval Augmented Generation (Graph-RAG)}

LLMs are statistical models that learn to predict the probability distribution of token sequences given input tokens. Advances in transformer architectures via the attention mechanism~\citep{vaswani2017attention} and large-scale pre-training~\citep{radford2018improving} have enabled LLMs to demonstrate impressive capabilities in understanding queries and generating more coherent responses. However, LLMs are not inherently aware of all domain knowledge up to their training date, nor are they guaranteed to have been exposed to every relevant piece of information during training. In particular, P\&ID diagrams and engineering design data are often confidential, meaning LLMs may not have had access to them during training. As a result, they may produce responses that appear fluent and coherent, but factually incorrect, a phenomenon often referred to as ‘hallucination’~\citep{dang2025survey}. Additionally, the finite context window of an LLM restricts the amount of external information it can effectively include during an inference~\citep{liu2024lost,kuratov2024search}. These limitations pose significant risks in safety-critical environments such as chemical-processing facilities, where decision-making demands high accuracy and reliability. 

Retrieval-augmented generation~(RAG)~\citep{lewis2020retrieval} addresses this limitation by enabling an LLM to retrieve and incorporate relevant information from external knowledge sources during inference. By grounding generation in specified domain data, RAG improves the accuracy, specificity, and factuality of model responses. The original RAG formulation is relatively simple: text documents are segmented into chunks and encoded as vectors, and a similarity-based retrieval module selects relevant content to provide as context for the model, a configuration often referred to as vector or naïve RAG~\citep{gao2023retrieval}. While this approach has demonstrated promising applicability across diverse domains, technical challenges remain, such as optimizing chunking strategies to manage retrieval latency~\citep{hasan2025engineering} and improving transparency and explainability, since retrieved information resides in high-dimensional embedding spaces.

GraphRAG extends traditional RAG by representing contextual knowledge as a graph of entities and their relationships. In GraphRAG, source documents are first transformed into a knowledge graph by identifying subjects and objects as nodes and their relationships as edges. This transformation can be achieved either through named entity recognition~(NER) and relation extraction using deep learning methods~\citep{li2020survey}, or by prompting LLMs to extract and summarize concepts (subject, object, and predicate) directly from text~\citep{edge2024local}. The resulting knowledge graph can be further semantically enriched and embedded in a vector space for indexing at a later stage. Once constructed, the knowledge graph can be queried using a variety of GraphRAG techniques, which broadly fall into three categories: (i) structured graph query languages (e.g., Cypher, GQL, Gremlin), (ii) vector-based retrieval using vector similarity search, and (iii) LLM-driven approaches in which the model either interprets the graph directly or generates graph queries, as well as combinations thereof. 

Recent GraphRAG approaches combine multiple retrieval methods. For example,~\citet{edge2024local} employs LLMs to summarize clusters of closely connected nodes into local semantic representations, then subsequently aggregates them into global semantic communities for later retrieval. \citet{sun2023think} introduces Think-on-Graph~(ToG), instead of simply using an LLM to query a knowledge graph~($\text{LLM} \oplus \text{KG}$), he use the native graph traversal algorithm, such as n−w GRAG (breadth-first) and n−d GRAG (depth-first), and LLMs as agents to interactively traverse and reason over the graph~($\text{LLM} \otimes \text{KG}$). \citet{ma2024think}, extend this work by implementing community detection to improve retrieval speed by guiding LLMs toward semantically coherent regions of the graph. Recent work also explores how LLMs can infer missing knowledge from incomplete graph structure, for instance, Graph Inspired Veracity Exploration~(GIVE) addresses retrieval challenges in GraphRAG systems~\citep{he2024give}.

GraphRAG is particularly interested in P\&ID analysis because P\&IDs are inherently graph-structured representations of process designs with symbols, making graph-based modeling both intuitive and natural. Indeed, several studies have already demonstrated the effectiveness of graph-based representations for P\&ID-related chemical process design and optimization tasks \citep{theisen2025graph, balhorn2025graph, alimin2025talking,  schulze2025rule}. Recent innovations in GraphRAG techniques also closely mirror how engineers interpret, trace, and abstract information from P\&IDs. Despite this strong conceptual alignment, no existing work has systematically evaluated GraphRAG approaches in the P\&ID and broader process engineering domains.

\subsection{Agentic Systems}
\label{subsec: Agentic Systems}

The deployment of LLMs typically spans a spectrum of complexity, from prompt engineering to agentic workflows. In prompt engineering, strategies such as zero-shot, one-shot, and few-shot prompting are commonly used. In zero-shot prompting, the LLM is given no examples; in one-shot prompting, it receives a single example; and in few-shot prompting, it is provided with a small set of examples to guide task completion. These approaches rely entirely on the model's pre-trained internal weights to generate responses from instructions or examples, without access to external data~\citep{brown2020language}. Techniques such as chain-of-thought prompting~\citep {wei2022chain} can further improve reasoning by encouraging the model to generate intermediate reasoning steps. While effective for general-purpose tasks, these "stateless" methods often underperform in specialized engineering domains because the model lacks access to proprietary information and cannot verify its outputs, leading to frequent hallucinations. 

\begin{figure}
    \centering
    \includegraphics[width=0.7\linewidth]{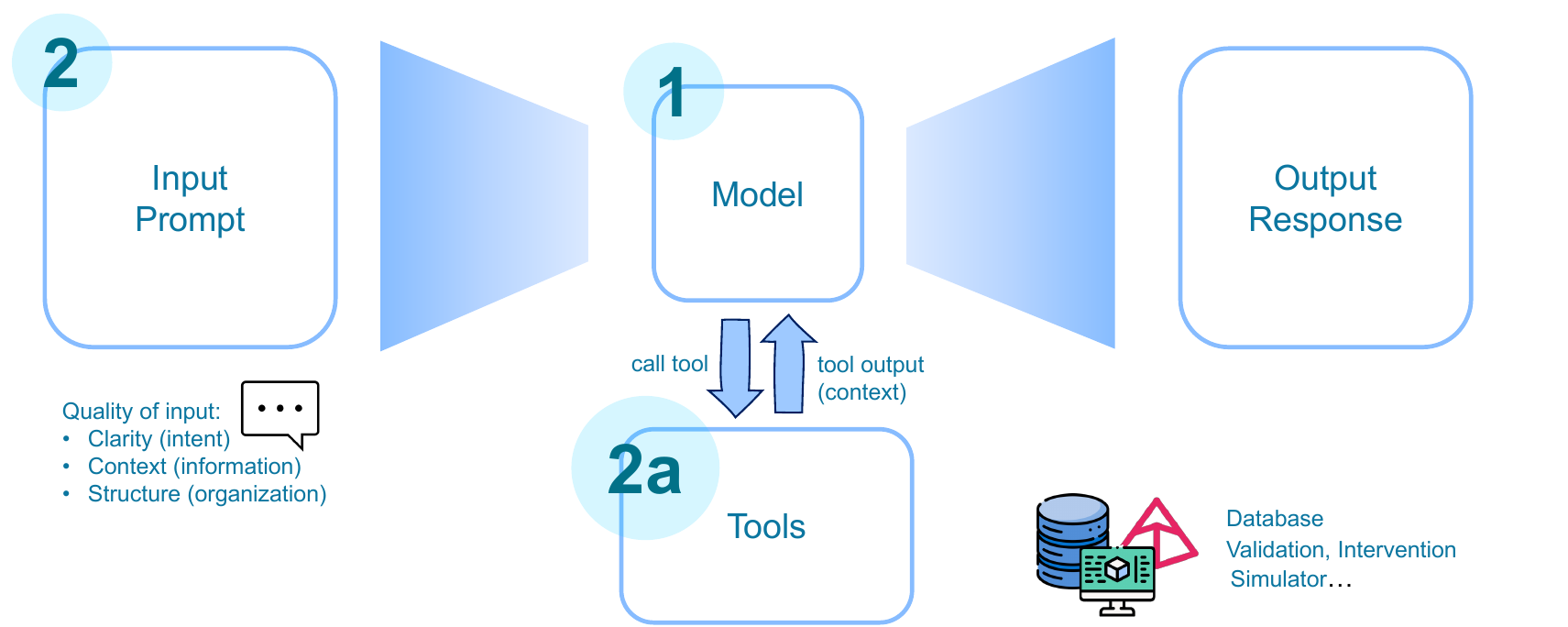}
    \caption{Agentic workflow illustrating how LLMs interact with external tools. The input prompt (2) provides intent, context, and structured information to the model (1). When needed, the model invokes tools (2a) to obtain additional information or perform tasks, and the tool's output serves as context that improves the final response.}
    \label{fig: agent}
\end{figure}

To address these limitations, recent research has shifted towards agentic workflows. Fundamentally, an agent is an intelligent system capable of perceiving its environment, planning actions, and executing tasks to achieve specific goals; this planning process requires understanding of context, multi-step reasoning, and adaptive decision-making~\citep{russell1995modern, ghallab2004automated}. In the context of LLMs, this is often formalized as the ReAct~(Reason and Act) workflow, where the model interleaves reasoning traces with executable actions~\citep{yao2022react}. As illustrated in Figure~\ref{fig: agent}, an agentic workflow operates through a dynamic interaction loop initiated by a user prompt. The prompt conveys the user’s intent and positions the LLM as the central reasoning engine. Rather than producing an immediate static response, the model first evaluates whether external information is required. If so, it invokes provided tools to interact with the environment, for example, querying a P\&ID knowledge graph database, performing web searches, or executing calculation modules. The outputs from these tools are then incorporated into the model’s context, allowing the agent to iteratively refine its understanding and generate responses grounded in validated engineering data rather than purely probabilistic inference.

In process engineering, the task can be complex and may require multiple steps to solve. Consequently, recent approaches to solving this problem have focused on implementing a multi-agent framework for chemical engineering~\citep{rupprechtMultiagentSystemsChemical2025}. Rather than attempting to solve a complex engineering problem in a single agent, agentic systems decompose it into smaller, well-defined subtasks, such as retrieving equipment specifications, accessing chemical safety data sheets (MSDS), chemical properties~\citep{bran2023chemcrow}, or performing vapor–liquid equilibrium (VLE) calculations. A supervisor agent can be designed to plan these subtasks and delegate them to multiple specialized agents equipped with distinct tools. This multi-agent coordination potentially reduces the likelihood of hallucinated yet plausible answers and enables iterative refinement of intermediate results when solving complex engineering problems. Furthermore, agentic workflows improve system traceability, as errors can be localized to specific reasoning steps or tool interactions, facilitating debugging and systematic improvement.

\section{Methods: ChatP\&ID}
\label{section: ChatP&ID}

\begin{figure}
    \centering
    \includegraphics[width=0.8\linewidth]{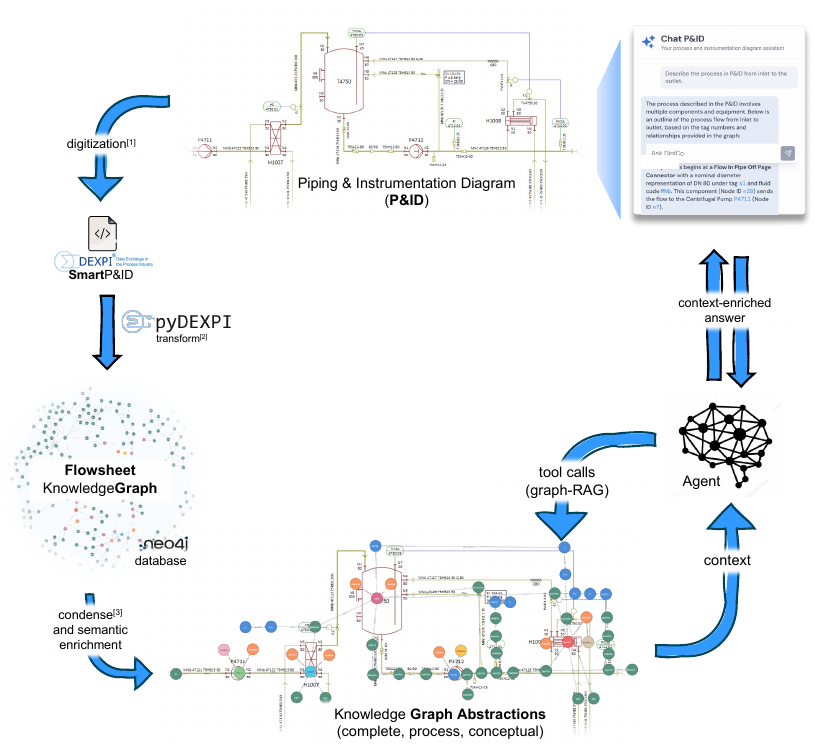}
    \caption{ChatP\&ID Workflow. The ChatP\&ID takes a flowsheet knowledge graph as its input. The knowledge graph can be generated by digitizing a PDF or image~\citep{theisen2023digitization} and exporting it to a DEXPI-conformant smart P\&ID.  The DEXPI P\&ID then transformed into a knowledge graph using pyDEXPI~\citep{Goldstein2025}. Using the pyDEXPI object, each flowsheet element is mapped one-to-one to a node in the knowledge graph. From this knowledge graph, multiple abstraction layers are created, from a complete graph to process-level and conceptual-level. GraphRAG methods: ContextRAG, VectorRAG, PathRAG, or CypherRAG, are implemented as tools to retrieve relevant information from a knowledge graph. The LLMs are integrated into the workflow as an agent that determines the appropriate retrieval method to invoke. Finally, a context-aware response is generated to answer user queries.}
    \label{fig: framework}
\end{figure}

This section presents our proposed workflow of ChatP\&ID, as illustrated in Figure~\ref{fig: framework}. The framework consists of three main components: (i) the agentic framework, (ii) the flowsheet knowledge graph generation, and (iii) the GraphRAG tools. In Section~\ref{subsec: Agentic framework}, we first describe the design and implementation of our agentic framework, including the underlying libraries, agent arrangement, and configuration of the chat interface. Next, Section~\ref{subsec: Flowsheet knowledge graphs} details the construction of the flowsheet knowledge graph, the abstraction process, and the associated semantic enrichment and embedding procedures. Finally, Section~\ref{subsec:GraphRAG} introduces the proposed GraphRAG methodology, which comprises ContextRAG, VectorRAG, PathRAG, and CypherRAG.

\subsection{Agentic framework}
\label{subsec: Agentic framework}

\begin{figure}
    \centering
    \includegraphics[width=0.7\linewidth]{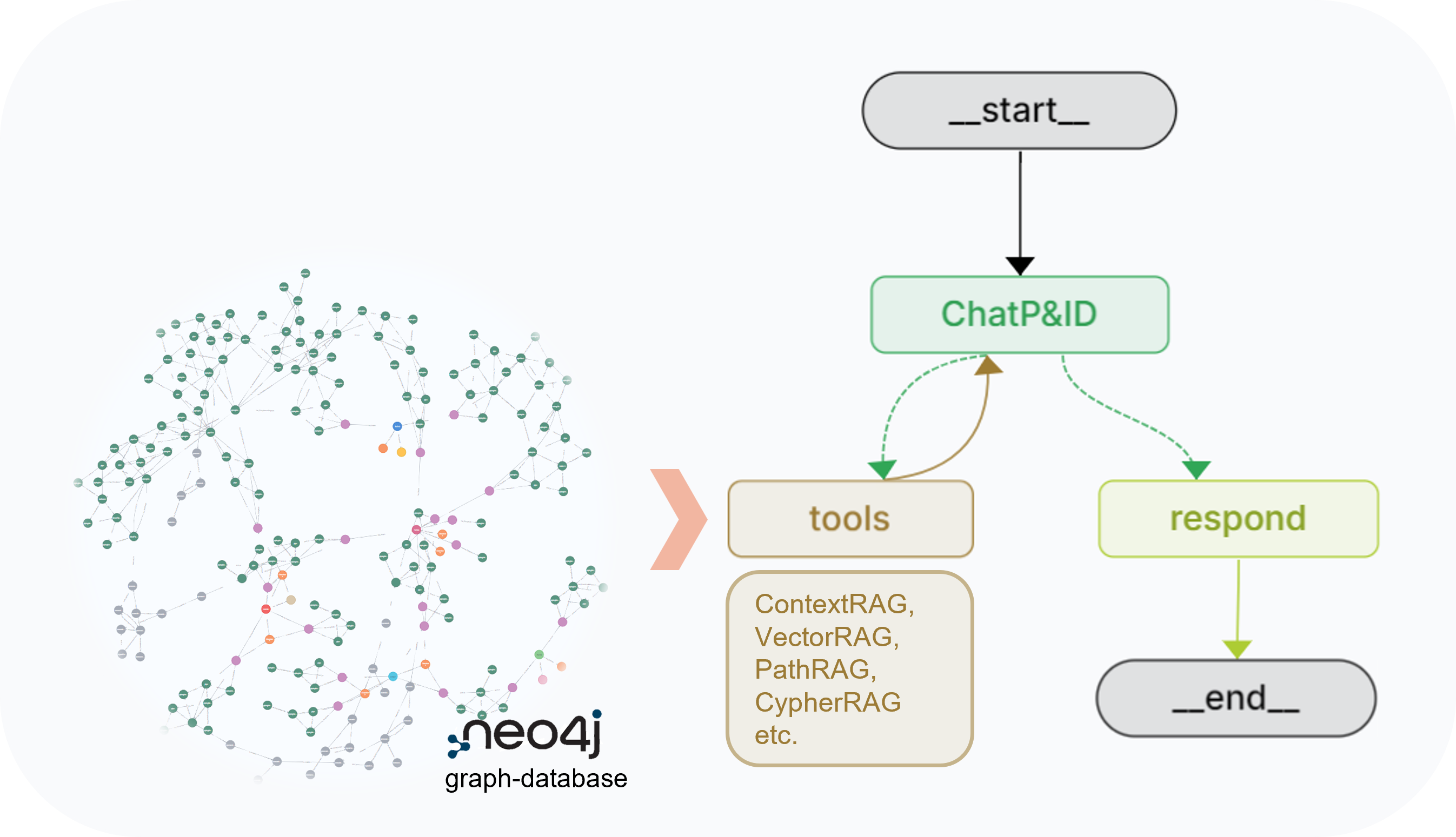}
    \caption{ChatP\&ID agentic workflow integrating graph database and graph-based retrieval method.}
    \label{fig:agent-workflow}
\end{figure}

At the core of ChatP\&ID is an LLM implemented as a reasoning engine within an agentic framework. We set up the agentic framework using the LangGraph library~\citep{langgraph2024} and configured it as illustrated in Figure~\ref{fig:agent-workflow}. When a user query is received, it is first routed to the ChatP\&ID agent, which determines whether a GraphRAG tool is needed and selects the appropriate information-retrieval method. To support context grounding, ChatP\&ID integrates a suite of graph-retrieval tools: ContextRAG, VectorRAG, PathRAG, and CypherRAG, which are described in Section~\ref{subsec:GraphRAG}. These tools query the Neo4j graph database to extract relevant nodes, edges, or sections of the knowledge graph and return them as textual information to the ChatP\&ID agent. Once a tool is selected, the ChatP\&ID agent also provides the necessary execution parameters for the tool, such as query mode (topology or graph) for ContextRAG~(Section~\ref{subsubsec:graph_abstraction_as_context}), index (local or global) for VectorRAG~(Section~\ref{subsubsec:vector_rag}), maximum traversal (depth and breadth) for PathRAG~(Section~\ref{subsubsec:path_rag}), or query for CypherRAG. The tool then executes the retrieval task and returns the results to the ChatP\&ID agent as additional context. The agent iteratively assesses whether the retrieved knowledge is sufficient to answer the user query. If additional context is required, the LLM may invoke additional GraphRAG methods until the information is sufficient or the predefined maximum number of tool calls is reached. The collected information is then aggregated and synthesized into a final, context-aware response.

To enable fluid multi-turn interactions, a memory module is integrated into the workflow, allowing conversation history to persist as contextual input across dialogue turns. This ensures continuity and enhances the consistency of reasoning across successive user interactions. Additionally, token streaming is employed to progressively display generated responses, improving user experience during longer reasoning or retrieval processes. Finally, a dedicated chat interface is developed to facilitate user interaction with ChatP\&ID. The interface streams tokens in real time, displays the conversation history, and provides visibility into the agent’s execution state, including tool invocation events and the corresponding retrieval results.

\subsection{Flowsheet knowledge graphs}
\label{subsec: Flowsheet knowledge graphs}
This section details the construction, abstraction, semantic enrichment, and embedding of a flowsheet knowledge graph derived from DEXPI P\&ID.

\subsubsection{Knowledge graph generation and abstraction}
\label{subsubsec:Knowledge graph generation and abstraction}

The knowledge graph is generated from DEXPI P\&ID data using our open-source package pyDEXPI \citep{Goldstein2025}, where each pyDEXPI instance is represented as a graph node, and the class attribute is represented as a graph relationship and property. The pyDEXPI data model defines three attribute types: compositional, reference, and data. In our representation, compositional attributes are modeled as \textit{has}-type relationships that form compositional edges between nodes. Reference attributes, such as \textit{send\_to}, \textit{control}, \textit{manipulate}, and other interaction-based relationships, are translated into relational edges connecting the corresponding nodes. Data attributes are preserved as node properties while maintaining the original naming conventions defined in pyDEXPI. Furthermore, the pyDEXPI class hierarchy is encoded as node labels.

For the development of a chat-based assistant, it is essential that the P\&ID representation incorporates comprehensive classification information (e.g., specific valve types such as globe, butterfly, needle, or gate valves) and detailed engineering specifications (e.g., fail-safe positions, nominal diameters, and material constraints). Given the broad scope of the proposed interface, the system must be able to retrieve and present all relevant P\&ID data. The LPG graph format is particularly well-suited for representing such attribute-rich engineering data, as it allows properties to be directly associated with nodes and edges. Therefore, in this work, we choose LPG as the underlying graph data model and host the graph in the Neo4j database.

In our previous work \citep{alimin2025talking}, we constructed a knowledge graph that provided both a one-to-one mapping from pyDEXPI and a high-level abstraction. This abstraction is necessary because the one-to-one mapping captures excessive detail that often diverges from how a process engineer conceptualizes a P\&ID. For instance, in the one-to-one mapping, a piping network system is modeled as a composition of segments, which in turn are compositions of individual components (e.g., valves, pipes, and fittings). This hierarchy yields a highly complex, convoluted graph structure. To address this, we previously condensed the graph to focus directly on piping components, aligning the representation with standard P\&ID interpretation. However, this reduction was performed as a single-step transformation from the complete graph to the high-level graph, thereby risking the loss of significant topological information.

Therefore, in this work, we implement a multi-step graph condensation process based on node classes (e.g., piping, equipment, and instrumentation) and refine the naming convention to provide more intuitive interpretations. Specifically, we construct three levels of graph abstraction: (1) the \textit{complete-level graph}, which provides a direct one-to-one mapping of all pyDEXPI entities and relationships; (2) the \textit{process-level graph}, which condenses the piping system from the complete-level graph; and (3) the \textit{conceptual-level graph}, which further condenses piping systems, instrumentation systems, and equipment from the process-level graph. For each abstraction, we apply: (i) pruning of domain-specific information and connection nodes, (ii) collapsing low-information nodes into their parent nodes, and (iii) removal of non-process-related node properties. The resulting graph is formally denoted as $G=(V, E)$, where each node $V$ contains data attributes and the corresponding pyDEXPI object, with node labels representing pyDEXPI classes, and edges $E$ encode compositional and reference relationships extracted from pyDEXPI.

\subsubsection{Knowledge graph semantic enrichment and embedding}
\label{subsec: Knowledge graph semantic enrichment and embedding}

In this section, we describe the process of enriching the knowledge graph generated in Section~\ref{subsubsec:Knowledge graph generation and abstraction} and projecting its nodes into a vector space. As discussed in Section~\ref{subsec: Graph-based Retrieval Augmented Generation (Graph-RAG)}, many GraphRAG techniques rely on vector-based retrieval to perform similarity searches over graph entities. Consequently, generating vector embeddings for each node is a prerequisite. These embeddings are critical for VectorRAG~(Section~\ref{subsubsec:vector_rag}), PathRAG~(Section~\ref{subsubsec:path_rag}), and other methods that utilize semantic similarity indexing to identify the nodes most relevant to a given query.
 
To set up these embeddings, we prompt LLMs to generate textual descriptions of each node that capture its semantics. We construct two types of semantic information for each node: (i) local semantics, which describe the node’s role relative to its immediate neighbors, reflecting its function within a localized subgraph, and (ii) global semantics, which describe the node’s role within the entire flowsheet, providing a high-level process perspective. We design this indexing strategy to support a coarse-to-fine retrieval process: global semantics enable the identification of specific nodes across the entire graph based on query similarity, while local semantics facilitate refinement within the identified neighborhood. This local and global approach has also been studied within the graphRAG system across different setups; for instance,~\citet{edge2024local} uses LLMs to generate (local) community summaries from graph nodes and edges. This community summary is subsequently aggregated for global reasoning, whereas \citet{shi2025deep} uses LLM-generated semantic representations for text nodes and embeds them into graph transformer models to capture both local and global graph structural information.

In our setup, both local and global semantics are generated by prompting an LLM; in this case, we use OpenAI GPT-4o to describe the node either in the context of its immediate neighbors or within the full graph (as graphml). Recent studies indicate that frontier models from Anthropic and OpenAI, such as Claude and GPT, demonstrate high proficiency in interpreting code and structured data formats, such as XML~\citep{nam2024using, derose2024can}. Hence, selected GPT-4o (OpenAI frontier model) based on our previous study~\citep{alimin2025talking}, which demonstrated its high accuracy in describing flowsheet information regardless of graph complexity. Furthermore, it maintains a reasonable token cost, which is critical given that this algorithm is executed for every node in the graph. We implement an algorithm that collects node information, local context, and global context, packages them into a single textual input along with the prompt, and passes it to the LLM. The full semantic enrichment procedure is detailed in Algorithm~\ref{alg:semantic_enrichment}, while the specific prompts and examples of the resulting semantic descriptions are provided in Appendix~\ref{app:semantic_generation}.

Once the semantic information is generated, the textual descriptions are converted into vector representations using embedding models. In this work, we use the Voyage embedding model, specifically voyage-3.5-lite, which has demonstrated high accuracy among publicly available models~\citep{butler2025massive, ponwitayarat2025sea, goel2025sage}. The model generates 1024-dimensional vectors, which serve as node embedding in downstream tasks. These embeddings enable similarity searches based on either local or global semantics in VectorRAG~(Section~\ref{subsubsec:vector_rag}) and PathRAG~(Section~\ref{subsubsec:path_rag}).

\begin{algorithm}
\caption{Semantic Enrichment and Embedding}
\label{alg:semantic_enrichment}
\begin{algorithmic}[1]
\Procedure{SemanticEnrichmentEmbedding}{$embeddingModel$}
    \State \textbf{Input:}   $embeddingModel$
    \State \textbf{Output:}  $V.embedding$ \Comment{Semantic embeddings for each node }
    \Statex
    \State Initialize: $embeddingModel$, $neo4jDriver$, $LLM$
    \Statex \Comment{Step 1: Retrieve all relevant nodes from graph}
    \State $nodes \gets$ \Call{GetAllNodes}{$graph$}
    \Statex \Comment{Step 2: Generate global and local semantics}
    \For{each $n \in nodes$}
        \State $global\_semantic \gets$ \Call{LLM.GenerateGlobalSemantic}{$n.labels$, $n.properties$, $graph$}
        \State $neighbors \gets \Call{GetNeighbors}{n}$
        \State $local\_semantic \gets$ \Call{LLM.GenerateLocalSemantic}{$n$, $neighbors$, $relationships(n)$}
        \State $n.semantic \gets (global\_semantic, local\_semantic)$
    \EndFor
    \Statex \Comment{Step 3: Encode semantics into vector embeddings using provided model}
    \For{each $n \in nodes$}
        \State $n.embedding \gets$ \Call{embeddingModel.Encode}{$n.semantic$}
    \EndFor

    \State \Return nodes
\EndProcedure
\end{algorithmic}
\end{algorithm}

\subsection{GraphRAG for P\&ID}
\label{subsec:GraphRAG}

To enable information retrieval from P\&ID knowledge graph during question answering, we propose four GraphRAG tools: ContextRAG, VectorRAG, PathRAG, and CypherRAG. These tools implement graph-based retrieval techniques, including LLM reads, vector search, and Cypher queries, or combinations thereof. They are implemented as LLM tools that can be invoked directly by the LLM during execution. When called, the tool executes the query algorithm on the Neo4j database and returns the results as textual context. The output can vary depending on the query and the tool, ranging from a single node or a list of nodes to a path or a section of the knowledge graph.

\subsubsection{ContextRAG}
\label{subsubsec:graph_abstraction_as_context}

ContextRAG is designed to extract a condensed, noise-free textual context from the knowledge graph. As background, LLMs have demonstrated significant capabilities for interpreting structured data, including code and native graph formats such as XML~\citep{nam2024using, alimin2025talking}. Frontier models, such as GPT-4 and Claude, are particularly effective at analyzing XML schemas, inferring relationships from markup, and identifying structural patterns from trees. However, these capabilities degrade significantly with input length. When processing large XML files, LLMs often encounter stability issues, such as losing track of global instructions, applying inconsistent reasoning, or failing to correctly associate related yet distant elements within the file~\citep{derose2024can}.  

In this tool, the Neo4j knowledge graph is serialized to a GraphML file, an XML-based format that explicitly encodes the graph topology, including nodes, edges, and attributes, as an XML tree. However, GraphML generated from pyDEXPI exports contains extensive metadata irrelevant to semantic reasoning, which increases noise and inflates token usage. To address this, in ContextRAG, we employ a filtering mechanism that retains only essential process structures and semantics. By stripping away extraneous attributes and focusing solely on flowsheet-relevant data, the system ensures that the retrieved context is both semantically dense and token-efficient.

ContextRAG proceeds in three stages. First, the target graph is exported from the database. Second, the raw GraphML data is processed to remove non-semantic artifacts, such as internal identifiers, URIs, and vector embeddings. Finally, the graph is reconstructed according to a agent-selected simplification mode: (i) \textit{Graph mode}, which preserves key node and edge attributes for detailed contextual reasoning by retaining essential P\&ID information from each component, including labels, tag number, and design specifications (e.g., design pressure, temperature ratings) following the pyDEXPI data attribute schema, alongside edge labels defining relationship types; meanwhile (ii) \textit{Topology mode} retains only node labels and connectivity for a lightweight structural overview. This approach yields a refined graph context that leverages the LLM's ability to parse structured XML to mitigate the stability issues arising from excessive length and noise.

\subsubsection{VectorRAG}
\label{subsubsec:vector_rag}

VectorRAG is designed to retrieve nodes through semantic similarity search. This tool performs a vector search over the global embeddings of the knowledge graph (detailed in Section~\ref{subsec: Knowledge graph semantic enrichment and embedding}) to identify nodes most relevant to the user's query. During execution, either the original user query or an LLM-refined version is transformed into a vector embedding using the Voyage-3.5-lite model, the same embedding model employed during the knowledge graph enrichment phase. This model projects the query into a 1024-dimensional vector space and then queries the pre-indexed node embeddings stored in the vector database. A cosine similarity search is performed to rank nodes by their semantic proximity to the query. Specifically, the similarity score $s_i$ between the query embedding $\mathbf{v}_q$ and a node embedding $\mathbf{v}_i$ is computed as:

\begin{equation}
\label{eq:semantic_similiarity}
s_i = \cos(\mathbf{v}_q, \mathbf{v}_i) = 
\frac{\mathbf{v}_q \cdot \mathbf{v}_i}{\|\mathbf{v}_q\| \, \|\mathbf{v}_i\|}
\end{equation}

The top-$k$ matched nodes, along with their enriched semantic content, are returned as contextual knowledge. The execution workflow is presented in Algorithm~\ref {alg:vectorrag}, which covers vector index initialization, similarity search over embeddings, and attribute filtering. By default, VectorRAG operates using global semantic embeddings. Local semantic embeddings are used when VectorRAG is invoked within PathRAG to identify semantically related neighboring nodes. This approach enables the integration of both global and local semantic representations for retrieval in downstream tasks.

\begin{algorithm}
\caption{VectorRAG: Semantic node retrieval}
\label{alg:vectorrag}
\begin{algorithmic}[1]
\Procedure{VectorRAG}{$indexName, query, topK$}
    \State \textbf{Input:} $indexName \in \{\text{``global\_semantic\_index''}, \text{``local\_semantic\_index''}\}$ \Comment{Vector index for embeddings}
    \State \textbf{Input:} $query$ \Comment{User query or question}
    \State \textbf{Input:} $topK$ (optional) \Comment{Max number of nodes to retrieve}
    \State \textbf{Output:} $results$ \Comment{List of relevant nodes with scores}
    \Statex
    \State Initialize: $embeddingModel$, $neo4jDriver$
    \State $retriever \gets$ VectorRetriever($indexName$, $embeddingModel$, $neo4jDriver$)
    \State $ranked \gets$ retriever.search($query$, $topK$)
    \State $results \gets \emptyset$
    \Statex

    \For{\textbf{each} $item$ \textbf{in} $ranked.items$}
        \State $content \gets$ ParseNodeContent($item.content$)
        \State $results \gets results \cup \big\{$
        \State score: $item.metadata.score$, 
            \State elementId: $item.metadata.id$, 
            \State nodeLabels: $item.metadata.nodeLabels$,
            \State content: $content$
            \State $\big\}$
    \EndFor

    \State \Return $results$
\EndProcedure
\end{algorithmic}
\end{algorithm}

\subsubsection{PathRAG}
\label{subsubsec:path_rag}
In PathRAG, we extend the VectorRAG framework by integrating path-based exploration into the retrieval process. This methodology mimics the cognitive workflow of process engineers, who typically analyze P\&IDs by first locating a specific component or area of interest and then tracing the connected process lines to establish context. Also, answering flowsheet queries often requires an understanding of these local topological relationships rather than finding specific information. For instance, to determine the isolation procedure for a vessel, an engineer must first identify the equipment and then trace the upstream and downstream paths to locate the corresponding isolation valves. Similarly, verifying a fluid transfer operation requires starting at a pump and traversing the discharge piping to confirm it connects to the intended tank. PathRAG is designed to operate on the 'locate-and-trace' strategy, enabling the system to derive answers from both the semantic identity of nodes and their structural connectivity.

\begin{figure}
    \centering
    \includegraphics[width=0.6\linewidth]{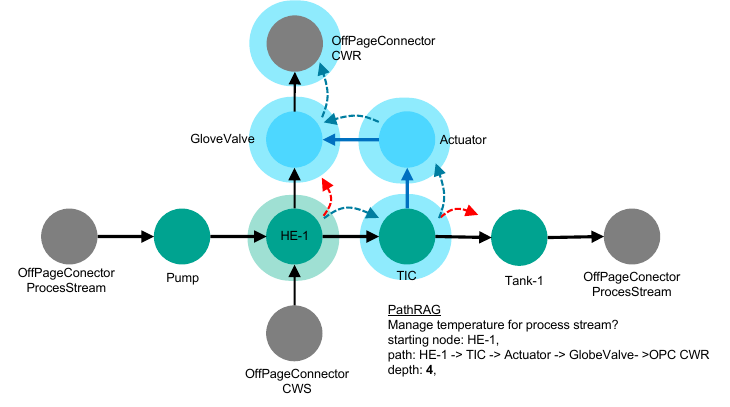}
    \caption{Illustration of the path exploration algorithm. To answer the query “how to control process stream temperature,” VectorRAG is first applied to the global embedding space to identify a suitable starting node. The Heat Exchanger~(HE) is selected as the starting point, as it is the semantically closest node to the query. From this node, two candidate paths exist: (i) a direct connection to the Globe Valve, or (ii) a path through the Temperature Indicating Controller~(TIC). VectorRAG is then applied to the local embedding of neighboring nodes to resolve this choice, selecting the TIC. The exploration proceeds through the Actuator and Globe Valve, and terminates at the Off-Page Connector~(OPC) for Cooling Water Return~(CWR). The resulting path is returned to the LLM to support the generation of the final answer.}
    \label{fig: path exploration}
\end{figure}

PathRAG incrementally traverses the graph along multiple paths, accumulating contextual knowledge at each step. The approach combines global and local semantic search to ensure that both relevant starting points and intermediate nodes are considered during query resolution. The procedure begins by embedding the user query and performing a global vector search to identify semantically relevant starting nodes. Each starting node serves as the root of a path exploration process, which is limited by a maximum depth and maximum breadth parameter to control computational complexity. The procedure is presented as in Algorithm~\ref{alg:pathrag}, where at each step along a path, we do (1) accumulates context from all previously visited nodes along the path, (2) evaluates, via the LLM, whether the accumulated context suffices to answer the query, (3) generates a next-hop query to guide local semantic search for the most relevant unvisited neighbor, and (4) updates the path with the selected node and its relationships, as visualized in Figure~\ref{fig: path exploration}.

\begin{algorithm}
\caption{PathRAG: Path exploration retrieval}
\label{alg:pathrag}
\begin{algorithmic}[1]
\Procedure{PathRAG}{$query, maxDepth, maxBreadth$}

    \State \textbf{Input:} $query$ \Comment{User query or question}
    \State \textbf{Input:} $maxDepth$ \Comment{Maximum number of hops}
    \State \textbf{Input:} $maxBreadth$ \Comment{Maximum parallel starting paths}
    \State \textbf{Output:} $result$ \Comment{Answer with exploration path}
    \Statex
    \State Initialize: $neo4jDriver$, $LLM$ 

    \Statex \Comment{Step 1: Global search for starting nodes}
    \State $startingNodes \gets$ VectorRAG(``global\_semantic\_index'', $query$, $maxBreadth$ )
    \If{$startingNodes = \emptyset$} \Return \textbf{NoAnswerFound} \EndIf
    
    \State $paths \gets \emptyset$
    
    \Statex \Comment{Step 2: Expand paths from each starting node}
    \For{\textbf{each} $nodeId$ in $startingNodes$}
        \State $path \gets [nodeId]$
        \State $visitedNodes \gets \{nodeId\}$
        \State $contexts[nodeId] \gets$ GetNodeContext($nodeId$)
        \State $accContext \gets$ BuildContext($contexts$)
    
        \If{EvaluateContext($query$, $accContext$).hasAnswer}
            \State append $(path, accContext, answer)$ to $paths$
            \State \textbf{continue}
        \EndIf
    
        \State $currentNode \gets nodeId$, $depth \gets 1$
        \While{$depth < maxDepth$}
            \State $neighbors \gets$ GetNeighbors($currentNode$) \textbf{excluding} $visitedNodes$
            \If{$neighbors = \emptyset$} \textbf{break} \EndIf
    
            \State $nextNode \gets$ VectorRAG(``local\_semantic\_index'', LLM.NextHopQuery($query$, $accContext$))
            \If{$nextNode = \text{None}$} \textbf{break} \EndIf
    
            \State append $nextNode$ to $path$
            \State $visitedNodes \gets visitedNodes \cup \{nextNode\}$
            \State $contexts[nextNode] \gets$ GetNodeContext($nextNode$)
            \State $accContext \gets$ BuildContext($contexts$)
    
            \If{LLM.EvaluateContext($query$, $accContext$).hasAnswer}
                \State append $(path, accContext, answer)$ to $paths$
                \State \textbf{break}
            \EndIf
    
            \State $currentNode \gets nextNode$, $depth \gets depth + 1$
        \EndWhile
        \State append $(path, accContext, answer)$ to $paths$
    \EndFor
    
    \Statex \Comment{Step 3: Select best answer}
    \State $result \gets$ LLM.SelectBestAnswer($paths$)
    \State \Return $result$

\EndProcedure
\end{algorithmic}
\end{algorithm}

This process continues until an answer is found, the maximum path depth is reached, or no unvisited neighbors remain. Multiple paths are explored in parallel from different starting nodes to enhance computational performance and robustness. Once all paths have been explored, the final answer is determined by the presence of relevant information along each path. If no path contains relevant information, the tool returns None. If at least one path contains relevant information, the tool returns the corresponding path(s), and the final answer is constructed by the LLM, which evaluates which path provides the most relevant information to answer the query.

\subsubsection{CypherRAG}
\label{subsubsec:cypher_rag}

CypherRAG leverages an LLM to translate natural language questions into executable Cypher queries for the Neo4j database. This approach is designed to enhance safety, as the database engine acts as a validator; any malformed or hallucinated syntax is rejected during execution rather than yielding incorrect data. Central to this process is the injection of the graph schema into the LLM's context window. Since the knowledge graph is constructed using pyDEXPI, this schema is intrinsically tied to the pyDEXPI data model. The schema is provided to the LLM as structural context alongside the user's query, enabling the model to map natural language intent to specific node labels and relationship types. Our implementation wraps the Neo4j Python library for Cypher-based question answering \citep{langchain_neo4j_cypher_docs}. The overall workflow is summarized in Algorithm~\ref{alg:cypherrag}. 

\begin{algorithm}
\caption{CypherRAG: Generates and executes Cypher for grounded retrieval}
\label{alg:cypherrag}
\begin{algorithmic}[1]
\Procedure{CypherRAG}{$query$}
    \State \textbf{Input:} $query$ \Comment{User query or question}
    \State \textbf{Output:} $answer$, $cypher$ \Comment{Grounded answer and executed query}
    \Statex
    \State Initialize: $neo4jDriver$, $LLM$
    \Statex
    \State $schema \gets$ GetGraphSchema()
    \Comment{Provide graph schema as context for Cypher generation}

    \State $cypher \gets$ LLM.GenerateCypher($query, schema$)
    \Comment{LLM translates question and schema into Cypher}

    \Statex
    \State $context \gets$ ExecuteCypher($cypher$)
    \Comment{Run Cypher query on Neo4j for factual grounding}

    \State $answer \gets$ LLM.Answer($query, context$)
    \Comment{LLM generates answer constrained to retrieved context}

    \Statex
    \State \Return $\{answer, cypher\}$
\EndProcedure
\end{algorithmic}
\end{algorithm}

\section{Illustrative case study}
\label{sec: Illustrative case study}

This section presents an illustrative case study used to evaluate the proposed framework, which extends to the test configuration, question sets, and evaluation procedures. 

\subsection{Configuration}
\label{subsec: Configuration}

\begin{figure}
    \centering
    \includegraphics[width=0.8\linewidth]{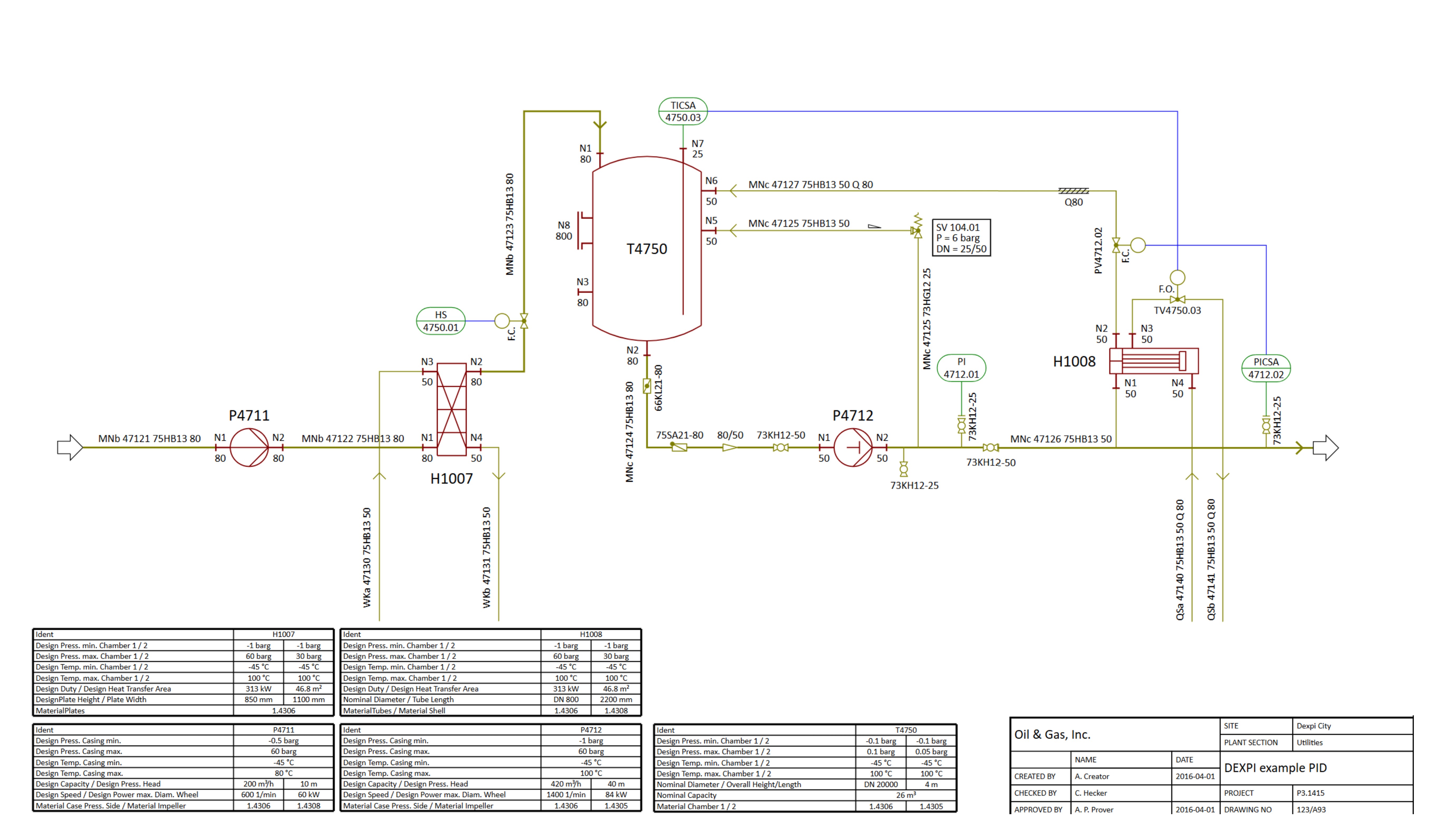}
    \caption{Sample of DEXPI P\&ID used in this case study~\citep{Theissen2021}.}
    \label{fig: dexpi p&id}
\end{figure}

The P\&ID used in this case study is the \textit{DEXPIEX01.xml} from~\citep{Theissen2021}, as shown in Figure~\ref{fig: dexpi p&id}. The flowsheet is processed through the pyDEXPI pipeline, converted into a knowledge graph, and stored in a graph database for use within the ChatP\&ID workflow. The evaluation spans the following dimensions: (1) Tasks: graph summarization, graph querying, path exploration, and knowledge inference; (2) Flowsheet representation: raw image, DEXPI Proteus file, and knowledge graphs at three abstraction levels (complete, process, and conceptual); (3) Tools: ContextRAG, VectorRAG, PathRAG, CypherRAG, read-image, and read-file; and (4) LLMs: OpenAI (GPT-4o, GPT-4o-mini, GPT-5, GPT-5-mini), Anthropic (Haiku 3.5, Sonnet 3.7, Sonnet 4, Opus 4.1), and Ollama-hosted models (Llama3.1:8B, Qwen3:4B, Qwen3:8B, Qwen3:14B, GPT-OSS:20B). Online models (OpenAI and Anthropic) are accessed via API, while offline models are executed locally using Ollama on a Mac Mini M4 with 24 GB RAM.

As noted earlier, direct interpretation of flowsheet images and DEXPI files by LLMs is included as a baseline method. The corresponding multimodal context is implemented by passing the P\&ID image as a base64-encoded string. Modern frontier LLMs already provide native multimodal capabilities, which enable image processing without external pre-processing. For image inputs, we ensure that equipment specifications are visible by providing high-resolution images (5,485 by 3,186 pixels, 330 DPI, 1,2MB) that include the associated specification tables, as illustrated in Figure~\ref{fig: dexpi p&id}. The Protheus context is implemented by passing the DEXPI Protheus file as a raw string to the LLM.  A tool limiter is added to limit the number of tool calls. Each tool can be called only once per test to evaluate its performance without iteration. For vector search, the maximum breadth is limited to 2 and the maximum depth to 3, considering the small size of our case P\&ID. Each benchmark configuration is executed twice to reduce stochastic variability in the reported results while keeping costs limited.

\subsection{Question-set}
\label{subsec: Question-set}

To evaluate the performance of GraphRAG and baseline methods for flowsheet information retrieval, we constructed a targeted flowsheet question–answer dataset consisting of 19 QA pairs. The questions are organized into four categories, each illustrated with a representative example. This dataset is designed to test a model’s ability to: (1) extract specific or multiple pieces of information, (2) comprehend the overall process structure, (3) explore graph topology and flow paths, and (4) combine concept in diagram information with inherent engineering information in the model. Representative samples from each category are detailed below:

\begin{enumerate}
    \item Graph query (10 questions)  \\
    \textit{Objective:} Evaluate the completeness of retrieving specific or multiple components and their attributes directly from the P\&ID. \\
    \textit{Sample Question:} (single query, 8 questions)  What is the design volumetric flow rate of pump P4712? \\
    \textit{Sample Question:} (multi query, 2 questions) List all valves in the P\&ID along with their specifications?

    \item Path exploration (5 questions) \\
    \textit{Objective:} Test the model’s ability to trace flow paths, identify alternative routes and isolation points, and infer operational or control logic from the graph. \\
    \textit{Sample Question:} If you need to isolate tank T4750 from all upstream equipment, which valves would you need to close?

    \item Knowledge inference (3 questions) \\
    \textit{Objective:} Test the model’s ability to derive insights from the diagram and connect it to the concept from internal information in the model. \\
    \textit{Sample Question:} What would be the effect if the heating fluid temperature in Heat Exchanger H1007 increased from the design temperature to 120 °C?

    \item Graph summarization (1 question) \\
    \textit{Objective:} Assess the model’s ability to interpret diagram components, relationships, and topology, and generate a coherent process narrative. \\
    \textit{Sample Question:} Based on the P\&ID, describe the process flow from the inlet to the final outlet, identifying major equipment, intermediate streams, and key control points in sequence.
    
\end{enumerate}

For each question, we provide a reference answer, to the best of our knowledge, derived from the P\&ID. This reference answer will help us compare with the LLM response. The set of our Q\&A is available in Appendix~\ref{app:qa set}.

\subsection{Evaluation}
\label{subsec:evaluation}

To assess the framework’s performance, we evaluate both the response accuracy and the computational cost. The response accuracy is assessed using two scoring approaches (1) Semantic similarity, where we compute the cosine similarity between the model-generated response and a reference answer in embedding space, as defined in Equation~\ref{eq:semantic_similiarity}. Both the model response and the reference answer are encoded into 1024-dimensional vectors using the Voyage-3.5-lite model. (2) LLM-as-a-Judge, where we employ LLM to assign a score to model responses relative to the reference answer based on predefined scoring rubrics. The evaluation is based on four criteria: relatedness, completeness, correctness, and coherence. The detailed rubric is shown in Table~\ref{table:rubrics}, which is adopted from~\citet{schulze2024empirical}. During evaluation, the LLM receives both the model output and the reference answer and is tasked with producing a score and a justification in accordance with the rubric. Scores range from 1 to 5, where higher scores indicate responses that are more aligned with the reference answer. For example, if the reference answer is “10 bar” and the model output is “5 bar,” the response would be considered incorrect but still related, partially complete, and coherent. To enable comparison with the semantic similarity metric, the rubric scores are linearly rescaled to the 0–1 range.

We use LLMs as judges due to the large number of evaluations required. Each configuration comprises 19 questions, each tested twice, and each test result is evaluated across five metrics. This yields more than 15,000 individual scoring points. Manually annotating this dataset would be labor-intensive; assuming a conservative rate of one minute per evaluation, the process would require approximately 250 person-hours (or 31 full working days).  Prior studies have demonstrated the effectiveness of LLMs as judges~\citep{tan2024judgebench, gu2024survey}. Notably, GPT-4 has been shown to achieve agreement rates above 80\% with human experts~\citep{zheng2023judging}. Consequently, we adopt GPT-5-mini, a lighter variant of the latest GPT-5 model from OpenAI, as our LLM judge for this study. However, it is important to note that LLM-as-judge may still result in hallucination when scoring; hence, we require the LLM to provide a step-by-step justification for every assigned score, and we analyze the reliability of these evaluations in Section~\ref{subsec:LLMs_vs_vector_similarity} by comparing their performance against semantic similarity metrics.

\begin{table}[t]
\centering
\caption{Rubric criteria for scientific answer evaluation for LLM (adapted from \citealp{schulze2024empirical}).}
\label{table:rubrics}
\setlength{\tabcolsep}{4pt}
\renewcommand{\arraystretch}{1.15}
\small

\begin{tabularx}{\textwidth}{p{2.8cm}>{\centering\arraybackslash}X>{\centering\arraybackslash}X>{\centering\arraybackslash}X>{\centering\arraybackslash}X>{\centering\arraybackslash}X}
\toprule
\textbf{Criterion} & \textbf{1} & \textbf{2} & \textbf{3} & \textbf{4} & \textbf{5} \\
\midrule
Relatedness 
& Not related at all 
& Mostly not related 
& Partly related 
& Mostly related 
& Completely related \\

Completeness 
& Incomplete, key details missing 
& Incomplete, insufficient details 
& Complete, insufficient details
& Complete, with most details
& Complete and detailed answer \\

Correctness 
& Completely incorrect 
& Mostly incorrect 
& Partly correct 
& Mostly correct 
& Completely correct \\

Coherence 
& Unable to draw conclusions from relevant knowledge
& Barely able to draw conclusions from established knowledge
& Can draw conclusions with some difficulty 
& Can independently draw conclusions based on established knowledge
& Can independently draw correct conclusions based on state-of-the-art knowledge \\

\bottomrule
\end{tabularx}
\end{table}

The computational cost was assessed to get the overview of (i) execution time (seconds), which shows the latency between query and last token from final response, (ii) cost (\$), the associated cost to answer one question, reflecting the cumulative number of input and output tokens. Both answer accuracy and computational cost were evaluated using the LangSmith library~\citep{langsmith2024} using our developed evaluation module. The experiments were conducted in September 2025, and the reported costs correspond to that period. Token-based cost analysis is applicable only to online models.

\section{Results}
\label{sec:Results}

This section presents an overview of ChatP\&ID's performance across different configurations and evaluation methods. We begin by comparing two evaluation approaches,  LLM-as-judge and semantic similarity, which will be the basis for interpreting LLM performance in the Results and Discussion sections. Then, we provide a high-level overview of the results, followed by a detailed analysis of how each configuration affects performance. Specifically, we examine the impact of flowsheet representations and graph abstraction levels, LLM selection, and GraphRAG tools across different task types.

\subsection{Contrasting LLM-based scoring with semantic similarity}
\label{subsec:LLMs_vs_vector_similarity}

\begin{figure}[ht!]
\centering
\begin{subfigure}[b]{0.48\linewidth}
    \centering
    \includegraphics[width=\linewidth]{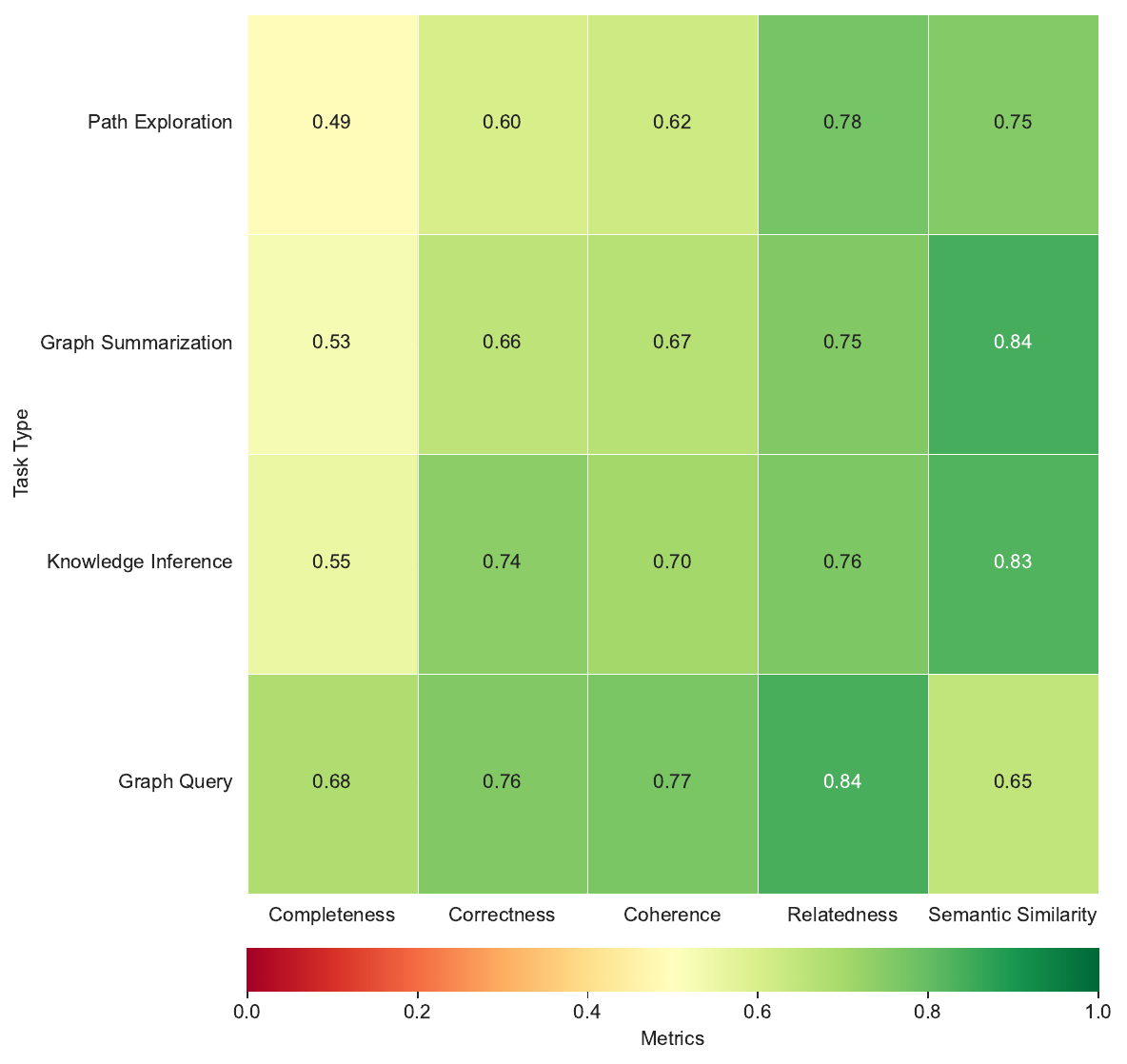}
    \label{fig:tools_metrics}
\end{subfigure}
\hfill
\begin{subfigure}[b]{0.45\linewidth}
    \centering
    \includegraphics[width=\linewidth]{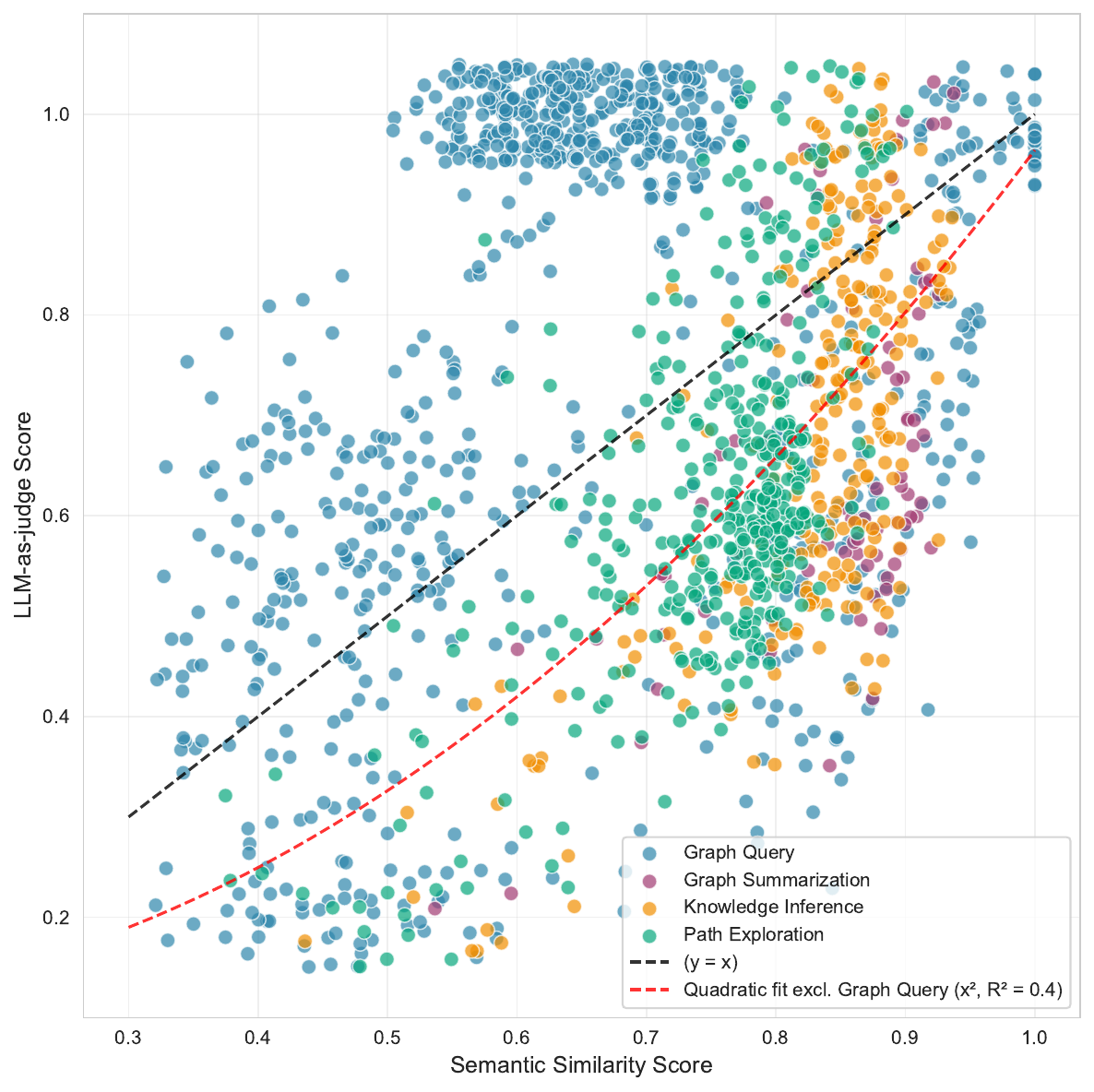}
    \label{fig:jitter}
\end{subfigure}
\caption{Evaluation of RAG strategies across task types: (left) Heatmap of response accuracy scores, (right) Relationship between average LLM-as-judge scores and semantic similarity. Small random noise (±0.05) is added to LLM scores for visibility.}
\label{fig:combined_metrics}
\end{figure}

As described in Section~\ref{subsec:evaluation}, we evaluate ChatP\&ID responses using two scoring approaches: (i) semantic similarity and (ii) LLM-as-judge. In this section, we compare these approaches based on their scoring results when evaluating LLM-generated responses. Figure~\ref{fig:combined_metrics} (left) shows the average score of each metric across different task types. Across the board, all metrics exhibit similar distributions, with a variation of approximately 15\%. Overall, the LLM-as-judge scores are fairly consistent across rubrics category. Completeness tends to score the lowest because responses must include all key information to receive a full score. In contrast, relatedness generally scores highest because it requires only that the response be relevant to the reference answer, not necessarily correct (e.g., providing an incorrect pressure value for an equipment specification may still be considered highly related).

Task-specific trends reveal a notable discrepancy between semantic similarity and LLM-as-judge scores for the graph query task (bottom right of the heatmap). While semantic similarity tends to score higher across other task types, the graph query result scores the lowest. This pattern is notable because, unlike other task types (where the reference answer may consist of a longer explanatory passage), the reference answer for graph queries is typically short and precise (e.g., “10 bar” or “4.0 m”). Consequently, graph queries demand exact factual matching rather than semantic or descriptive overlap. 

To further investigate, we examined the distribution of semantic similarity versus LLM-as-judge scores across tasks in Figure \ref{fig:combined_metrics} (right). The two scoring methods do not exhibit a linear correlation; instead, semantic similarity values are compressed toward the higher end. This relationship is sensitive to both the choice of embedding model (for semantic similarity) and the rubric scoring design (for LLM-as-judge). We also find that semantic similarity aligns more closely with average LLM-as-judge scores when graph query tasks are excluded. The scatter plot further shows that graph query dots deviate disproportionately from the overall trend, exhibiting substantially lower similarity scores than LLM-as-judge scores (in the middle top of the scatter plot).

\begin{figure}
    \centering
    \includegraphics[width=0.8\linewidth]{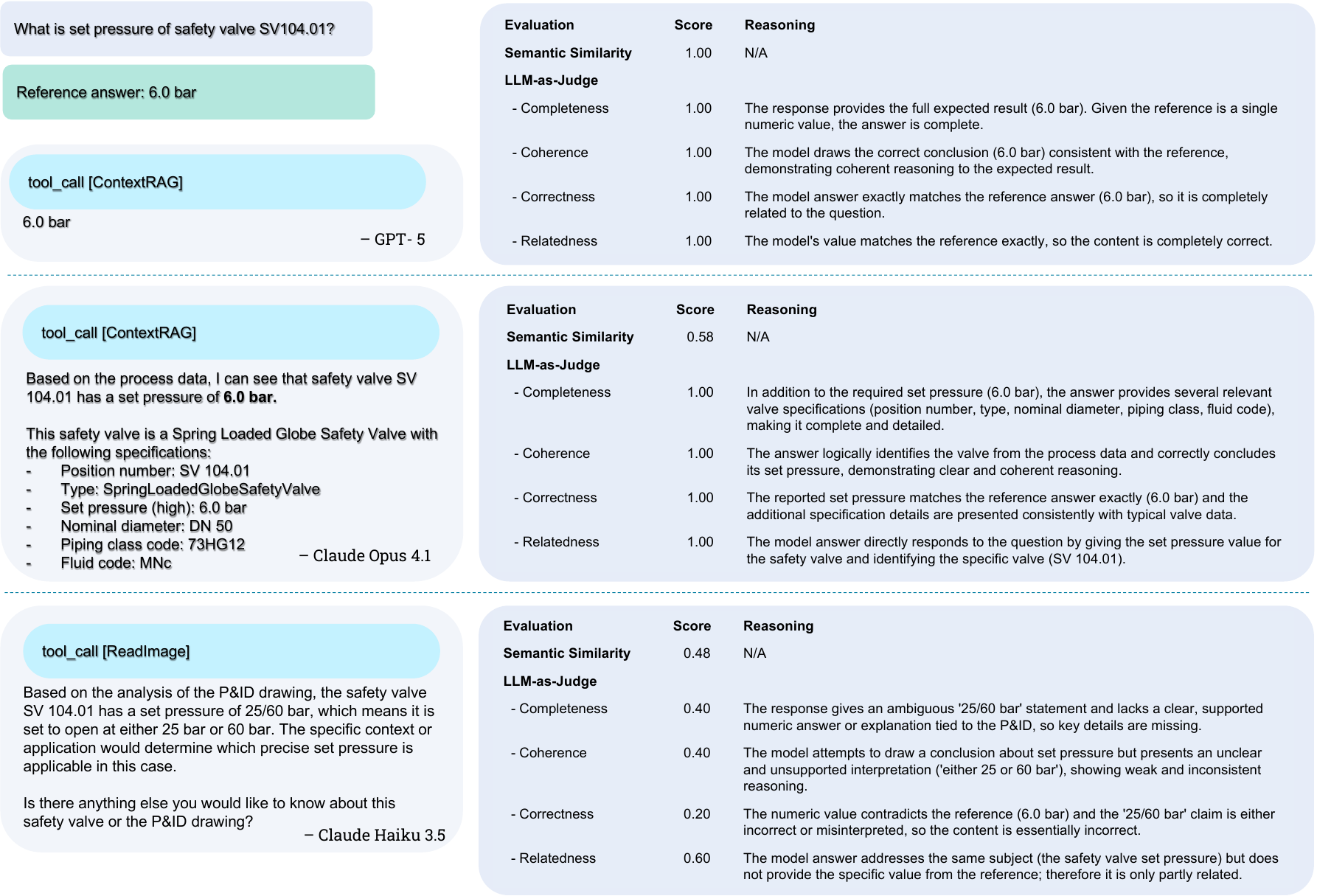}
    \caption{Comparison of semantic similarity and LLM-as-judge evaluations for a graph-query task retrieving the set pressure of a safety valve: (i) GPT-5 with ContextRAG, (ii) Claude Opus 4.1 with ContextRAG, and (iii) Claude Haiku 3.5 using read image.}
    \label{fig: Case PSV}
\end{figure}

We further examine this behavior with examples from our evaluation benchmark, shown in Figure~\ref{fig: Case PSV}. In this case, we compare responses from three LLMs: GPT-5, Claude Opus 4.1, and Claude Haiku 3.5, for finding the set pressure of the PSV. Both GPT-5 and Claude Opus 4.1 produce correct answers, differing primarily in the formulation of their responses. GPT-5 provides a concise and direct answer, "6.0 bar," for the PSV set pressure, achieving a perfect score (1.00) across all evaluation metrics, including semantic similarity. In contrast, Claude Opus 4.1 delivers a more detailed explanation. Although the answer and all supporting details are correct, its semantic similarity score is only 0.58, while the LLM-as-judge assigns a perfect score of 1.00. This discrepancy arises because the reference answer is brief, making it semantically distant from a longer, more descriptive response.

In contrast, Claude Haiku 3.5 provides a similarly detailed explanation but reports an incorrect numerical value. In this instance, the semantic similarity score is comparable to that of Claude Opus 4.1, yet the LLM-as-judge assigns a low correctness score (0.2) and explicitly explains its reasoning. The relatedness score remains relatively high (0.6), as the response addresses the same topic on the PSV set pressure but reports an incorrect value. These results demonstrate that semantic similarity reflects vector-space closeness to the reference answer rather than factual or logical correctness. Consequently, a response may be correct despite low semantic similarity, while a highly similar response may still be incorrect.

Moreover, LLM-as-judge evaluations can provide interpretability by generating explanations for their scores, offering insights that go beyond what semantic similarity alone can capture. Although semantic similarity is often used as a proxy for accuracy, our findings suggest that it primarily measures topical relevance rather than reasoning accuracy or correctness. Nevertheless, it remains a useful high-level indicator for semantic indexing. Based on these observations, we adopt LLM-as-judge as our primary evaluation method and report response accuracy by averaging the full set of accuracy metrics produced by the LLM-based evaluator.

\subsection{Performance overview}
\label{subsec: Overview}

\begin{figure}
    \centering
    \includegraphics[width=1.0\linewidth]{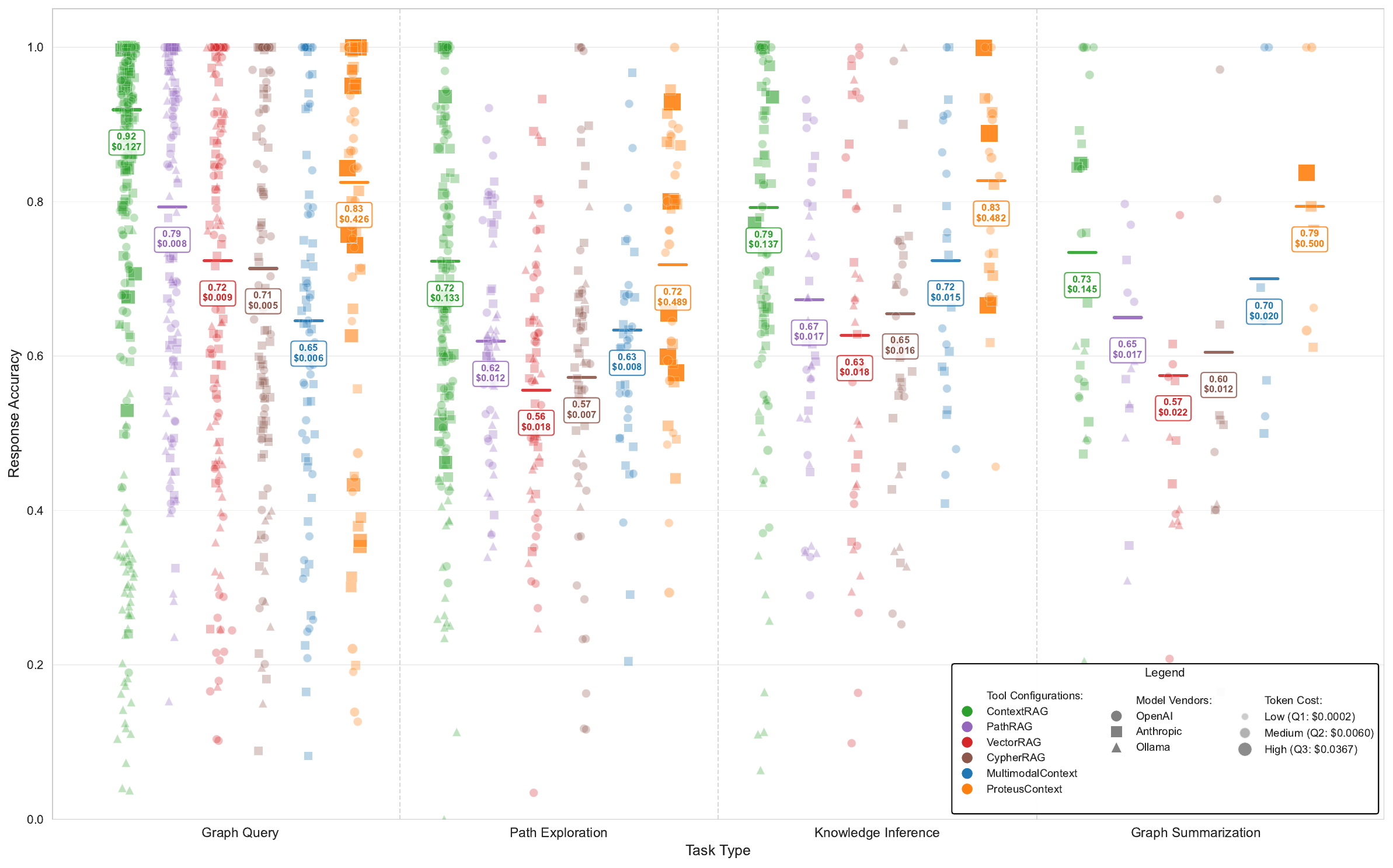}
    \caption{Performance overview of ChatP\&ID. The scatter plot groups results by four tasks: graph query, path exploration, knowledge inference, and graph summarization. The y-axis represents the response accuracy, calculated by averaging all LLM-as-judge metrics. Data points are color-coded by tool configuration and shaped by the LLM model host (OpenAI, Anthropic, Ollama). The size of each marker indicates the token cost per QA pair, with larger markers corresponding to higher costs. Horizontal bars with annotations display the mean accuracy score and mean cost for each tool configuration. The average of the accuracy score and cost excludes the Ollama-hosted model, as not all tasks can be solved by the model.}
    \label{fig:tools_task_performance}
\end{figure}

Figure~\ref{fig:tools_task_performance} summarizes GraphRAG performance across different LLM configurations and task types. The accuracy score is calculated by averaging all LLM-as-judge scores: correctness, relatedness, coherence, and completeness. Across all tasks, ContextRAG (green) achieved the highest average score (0.84), followed by Proteus context (orange) at 0.80. Notably, ContextRAG reached this performance at four times lower cost per task (\$0.13) compared to Proteus context (\$0.45).

The next-highest-performing methods were the vector-similarity-based approaches PathRAG (purple) and VectorRAG (red), with average accuracies of 0.72 and 0.71, respectively. These methods incurred very low token costs per task (\$0.010–\$0.015 per question), largely due to lower token usage enabled by vector similarity search. CypherRAG (brown) followed with an average accuracy score of 0.67.

Multimodal Context achieved the lowest overall average score (0.66) but also the lowest token cost per task (\$0.009). largely due to the lower cost of multimodal tokens (per tile) compared to text tokens, with a variable of image size and resolution. However, its performance was highly task-dependent: for graph queries and path exploration that require precise information retrieval (e.g., equipment specifications, flow tracing), accuracy dropped to 0.65 and 0.63, respectively. Finally, it is important to note that these average scores are influenced by the choice of LLM, as increasing flowsheet representation and tool complexity require more capable (larger) models, a factor we discuss further in the next section.

\subsection{Effect of flowsheet representation}
\label{subsec: Effect of flowsheet representation}
We evaluate LLMs' performance across different flowsheet representations and levels of graph abstraction, as summarized in Table~\ref{table:representation performance}. Overall, representing a P\&ID as a graph at any level of abstraction consistently improves LLM performance compared to directly providing the raw flowsheet image or Proteus file. For flagship models such as GPT-5, accuracy increases by approximately 5\% on average when using a conceptual-level graph as context rather than the Proteus file. In addition to accuracy gains, the conceptual graph substantially reduces computational cost by up to 85\% relative to the Proteus file, resulting in an absolute reduction of approximately \$0.15 per task. Smaller models show more significant improvements; for example, GPT-4o mini achieves approximately 7\% higher accuracy with conceptual graph representations compared to direct Proteus file input, a trend also observed in other online LLMs. Offline Ollama hosted models that we selected (Llama3.1, GPT-OSS, and Qwen3) were excluded from this evaluation due to limitations in processing multimodal images and the extended time required to parse Proteus files. The complete set of results is provided in the Appendix~\ref{app:complete results}.

\begin{table}[t]
\centering
\caption{Response accuracy and token cost per task across different knowledge graph abstractions and flowsheet representations. Tools ContextRAG is used to query from knowledge graphs, image read for flowsheet image, and file read for Proteus File. GPT-5 and GPT-4o-mini results are for comparison. The response accuracy is the average of LLM-as-judge score: completeness, coherence, correctness, and relatedness.}
\label{table:representation performance}
\setlength{\tabcolsep}{4pt}
\renewcommand{\arraystretch}{1.2}
\small
\begin{tabularx}{\textwidth}{p{3.0cm}>{\centering\arraybackslash}X>{\centering\arraybackslash}X>{\centering\arraybackslash}X>{\centering\arraybackslash}X>{\centering\arraybackslash}X>{\centering\arraybackslash}X}
\toprule
\textbf{Representation} & \textbf{Graph Query} (single query) & \textbf{Graph Query} (multi-query) & \textbf{Graph Summarization} & \textbf{Knowledge Inference} & \textbf{Path Exploration} & \textbf{Average} \\
\midrule
\multicolumn{7}{c}{\textbf{GPT-5}} \\
\midrule
Graph Abstractions & & & & & & \\
- Conceptual Level 
& 0.91 / \$0.011 
& 1.00 / \$0.033 
& 1.00 / \$0.064 
& 0.98 / \$0.055 
& 0.92 / \$0.027 
& 0.94 / \underline{\$0.027} \\
- Process Level 
& 0.93 / \$0.035 
& 0.96 / \$0.060 
& 1.00 / \$0.096 
& 0.98 / \$0.076 
& 0.87 / \$0.063 
& 0.93 / \$0.055 \\
- Complete Level 
& 0.95 / \$0.086 
& 0.75 / \$0.126 
& 1.00 / \$0.168 
& 0.98 / \$0.126 
& 0.82 / \$0.132 
& 0.90 / \$0.113 \\
\midrule
Flowsheet Image 
& 0.79 / \$0.002 
& 0.61 / \$0.022 
& 0.93 / \$0.027 
& 0.87 / \$0.020 
& 0.67 / \$0.008 
& 0.76 / \$0.001 \\
Proteus File 
& 0.89 / \$0.152 
& 0.74 / \$0.195 
& 1.00 / \$0.216 
& 0.98 / \$0.191 
& 0.87 / \$0.184 
& 0.89 / \underline{\$0.175} \\
\midrule
\multicolumn{7}{c}{\textbf{GPT-4o-mini}} \\
\midrule
Graph Abstractions & & & & & & \\
- Conceptual Level 
& 0.98 / \$0.001 
& 0.78 / \$0.001 
& 0.50 / \$0.001 
& 0.63 / \$0.001 
& 0.61 / \$0.001 
& \underline{0.78} / \$0.001 \\
- Process Level 
& 0.91 / \$0.004 
& 0.66 / \$0.004 
& 0.53 / \$0.004 
& 0.61 / \$0.003 
& 0.55 / \$0.004 
& 0.72 / \$0.004 \\
- Complete Level 
& 0.93 / \$0.010 
& 0.58 / \$0.010 
& 0.50 / \$0.010 
& 0.61 / \$0.007 
& 0.53 / \$0.010 
& 0.71 / \$0.010 \\
\midrule
Flowsheet Image 
& 0.57 / \$0.0001 
& 0.33 / \$0.0003 
& 0.68 / \$0.0004 
& 0.62 / \$0.0003 
& 0.55 / \$0.0001 
& 0.55 / \$0.0002 \\
Proteus File 
& 0.85 / \$0.018 
& 0.48 / \$0.018 
& 0.58 / \$0.019 
& 0.74 / \$0.012 
& 0.60 / \$0.018 
& \underline{0.71} / \$0.017 \\
\bottomrule
\end{tabularx}
\end{table}

When comparing levels of graph abstraction, flagship models such as GPT-5 leverage the complex relationships within complete graphs to achieve high accuracy on single-query tasks. However, other tasks achieve higher accuracy when using higher-level graph abstractions. Smaller models, such as GPT-4o mini, perform best at the highest level of abstraction, with accuracy decreasing as graph complexity increases across all task types. 

Directly providing raw flowsheet images yielded the lowest overall accuracy across the evaluation, averaging 0.76 for GPT-5 and 0.55 for GPT-4o-mini. The most significant performance degradation occurred in tasks requiring high technical precision, such as single- and multi-query graph retrieval. Conversely, vision-based methods performed best in high-level overview tasks, including graph summarization and knowledge inference. While this multimodal approach offers the lowest operational cost of all tested methods, its reliability varies significantly across task types. Consequently, while raw image processing may suffice for general descriptive tasks, it remains inadequate for complex industrial reasoning compared to the Proteus context or the GraphRAG method.

\subsection{LLM model evaluation and selection}
\label{subsec:LLM model evaluation and selection}

We assessed the performance of various LLMs across GraphRAG methods, multimodal context, and Proteus context. Our evaluation included models from Anthropic, OpenAI, and Ollama, tested across different tool configurations. For the GraphRAG implementation, we use conceptual-level knowledge graphs as the flowsheet graph representation; as discussed in Section~\ref{subsec: Effect of flowsheet representation}, this format consistently yields the highest accuracy across both large and small models, ensuring a fair comparison when varying model scale. Additionally, multimodal and Proteus contexts were evaluated for all models to serve as performance baselines. The results are summarized in Table~\ref{table:tool_model_performance_cost_3dec}.

Overall, larger models tend to achieve higher accuracy, though at the cost of greater computational resources. Online models from Anthropic and OpenAI exhibited an average task completion time of approximately 30 seconds, peaking at 90 seconds when using the PathRAG tool, due to the path exploration algorithm's longer execution time. While individual inference times are not detailed here (refer to the Appendix~\ref{app:complete results} for the full breakdown), they stand in contrast to locally hosted Ollama models. These open-source offline configurations required more processing time, with durations ranging from several minutes to upwards of 15 minutes per task. Note that these offline Ollama-hosted models were run on a Mac mini M4.

\begin{table}[t]
\centering
\caption{LLMs performance across GraphRAG tools, multimodal context, and Proteus context. Various models from Anthropic, OpenAI, and Ollama are tested. Anthropic and OpenAI are compared by the response accuracy and token cost per task. While the Ollama model is measured by response accuracy and the time needed to complete the task. The Ollama was run in a Mac mini M4 with 24GB RAM. The flowsheet is represented as a conceptual graph for GraphRAG, a flowsheet image for Multimodal Context, and a Proteus XML file for Proteus Context. The response accuracy is the average of LLM-as-judge score: completeness, coherence, correctness, and relatedness.}
\label{table:tool_model_performance_cost_3dec}
\setlength{\tabcolsep}{4pt}
\renewcommand{\arraystretch}{1.2}
\small
\begin{tabularx}{\textwidth}{p{3.5cm}>{\centering\arraybackslash}X>{\centering\arraybackslash}X>{\centering\arraybackslash}X>{\centering\arraybackslash}X>{\centering\arraybackslash}X>{\centering\arraybackslash}X}
\toprule
\textbf{Model} & \textbf{ContextRAG} & \textbf{VectorRAG} & \textbf{PathRAG} & \textbf{CypherRAG} & \textbf{Multimodal Context} & \textbf{Proteus Context} \\
\midrule
\multicolumn{7}{c}{\textbf{Anthropic} (response accuracy / cost per task)} \\
\midrule
Claude-3-5-Haiku 
& 0.82 / \$0.008 
& 0.71 / \$0.003 
& 0.72 / \$0.002 
& 0.54 / \$0.001 
& 0.59 / \$0.002 
& 0.76 / \$0.123 \\
Claude-3-7-Sonnet 
& 0.88 / \$0.032 
& 0.71 / \$0.015 
& 0.72 / \$0.009 
& 0.61 / \$0.005 
& 0.62 / \$0.007 
& 0.81 / \$0.464 \\
Claude-Sonnet-4 
& 0.87 / \$0.033 
& 0.71 / \$0.014 
& 0.72 / \$0.009 
& 0.69 / \$0.005 
& 0.68 / \$0.009 
& 0.81 / \$0.465 \\
Claude-Opus-4-1 
& \underline{0.85 / \$0.162} 
& 0.69 / \$0.068 
& 0.72 / \$0.045 
& 0.68 / \$0.024 
& 0.66 / \$0.038 
& 0.78 / \$2.077 \\
\midrule
\multicolumn{7}{c}{\textbf{OpenAI} (response accuracy/ cost per task)} \\
\midrule
GPT-4o-mini 
& 0.78 / \$0.001 
& 0.64 / \$0.0002 
& 0.64 / \$0.0002 
& 0.54 / \$0.0001 
& 0.55 / \$0.0002 
& 0.71 / \$0.017 \\
GPT-4o 
& 0.83 / \$0.027 
& 0.64 / \$0.003 
& 0.63 / \$0.003 
& 0.60 / \$0.002 
& 0.58 / \$0.002 
& 0.73 / \$0.286 \\
GPT-5-mini 
& \underline{0.91 / \$0.004} 
& 0.82 / \$0.002 
& 0.83 / \$0.002 
& 0.86 / \$0.002 
& 0.83 / \$0.002 
& 0.88 / \$0.034 \\
GPT-5 
& 0.94 / \$0.027 
& 0.79 / \$0.017 
& 0.80 / \$0.015 
& 0.79 / \$0.012 
& 0.76 / \$0.010 
& 0.89 / \$0.175 \\
\midrule
\multicolumn{7}{c}{\textbf{Ollama} (response accuracy / execution time per task)} \\
\midrule
LLaMA3.1:8B 
& 0.28 / 50s 
& 0.59 / 13s 
& 0.62 / 100s 
& * 
& ** 
& * \\
Qwen3:4B 
& 0.37 / 275s 
& 0.63 / 54s 
& 0.68 / 141s 
& * 
& ** 
& * \\
Qwen3:8B 
& 0.35 / 173s 
& 0.67 / 51s 
& 0.70 / 145s 
& 0.33 / 86s 
& ** 
& * \\
Qwen3:14B 
& 0.43 / 284s 
& 0.63 / 84s 
& 0.68 / 139s 
& 0.52 / 273s 
& ** 
& * \\
GPT-OSS:20B 
& 0.30 / 380s 
& 0.72 / 38s 
& 0.70 / 81s 
& 0.45 / 73s 
& ** 
& * \\
\bottomrule
\multicolumn{7}{l}{* Model exceeded the 900s execution time.} \\
\multicolumn{7}{l}{** The model did not have multimodal capability.} \\
\end{tabularx}
\end{table}

For online models, ContextRAG consistently yielded the highest accuracy across tasks (average 0.84) for both Anthropic and OpenAI models.  GPT-5-mini and GPT-5 demonstrated the highest accuracy, with GPT-5-mini achieving it at a fivefold lower cost, and reliably handling multimodal and Proteus contexts with an average accuracy of 0.85. For offline models, ContextRAG scores were lower for smaller models, reflecting the limited reasoning capability. Larger offline models achieved higher accuracy but incurred significantly longer execution times. 

Tools such as VectorRAG and PathRAG improved offline model accuracy, increasing the score from 20\% to 40\%. VectorRAG tasks were generally completed within one minute, while PathRAG tasks took around two minutes due to the need to traverse multiple nodes. CypherRAG tasks were computationally more complex, and only a few offline models could produce answers within the practical 15-minute time limit. Meanwhile, Proteus contexts presented additional challenges for offline models. In many cases, these offline models failed to produce results within the 15-minute cutoff. When answers were eventually produced, they were generally unsatisfactory and prone to hallucinations. The selected offline model does not support multimodal context.

\subsection{Tool performance comparison}
\label{subsec: Tool performance comparison}
We examine GraphRAG's performance across different query tasks and compare it with baseline methods. For this analysis, we selected GPT-5 mini as ChatP\&ID LLM, following the findings in Section~\ref{subsec:LLM model evaluation and selection}. This model was chosen for its consistently high performance across GraphRAG tools, multimodal, and Proteus contexts. By utilizing a model that performs reliably across all input formats, we ensure an equitable comparison of the underlying tool capabilities. The comparative results are summarized in Table~\ref{table:tool_performance_two_level}.

\begin{table}[t]
\centering
\caption{Tool performance across tasks: response accuracy, cost per task (\$), and execution time (s). The LLM examined is GPT-5-mini, the graph abstraction at the conceptual level is used to test GraphRAG, the image read tool is used to read Multimodal Context, and file read for Proteus Context. The response accuracy is the average of LLM-as-judge score: completeness, coherence, correctness, and relatedness.}
\label{table:tool_performance_two_level}
\setlength{\tabcolsep}{6pt}
\renewcommand{\arraystretch}{1.3}
\small
\begin{tabularx}{\textwidth}{p{3cm}>{\centering\arraybackslash}X>{\centering\arraybackslash}X>{\centering\arraybackslash}X>{\centering\arraybackslash}X>{\centering\arraybackslash}X>{\centering\arraybackslash}X}
\toprule
\textbf{Tool} & \textbf{Graph Query (single)} & \textbf{Graph Query (multi)} & \textbf{Graph Summarization} & \textbf{Knowledge Inference} & \textbf{Path Exploration} & \textbf{Average} \\
\midrule
\multicolumn{7}{c}{\textbf{response accuracy}} \\
\midrule
ContextRAG        & 0.89 & 0.95 & 1.00 & 0.98 & 0.88 & \underline{0.91} \\
VectorRAG         & 0.90 & 0.79 & 0.78 & 0.87 & 0.68 & 0.82 \\
PathRAG           & 0.87 & 0.65 & 0.88 & 0.88 & 0.79 & 0.83 \\
CypherRAG         & 0.88 & 0.81 & 0.93 & 0.86 & 0.84 & 0.86 \\
\midrule
Multimodal Context & 0.79 & 0.85 & 0.95 & 0.91 & 0.80 & 0.83 \\
Proteus Context   & 0.90 & 0.79 & 1.00 & 0.99 & 0.79 & 0.88 \\
\midrule
\multicolumn{7}{c}{\textbf{cost per task (\$)}} \\
\midrule
ContextRAG        & 0.0021 & 0.0058 & 0.0099 & 0.0078 & 0.0045 & 0.0044 \\
VectorRAG         & 0.0012 & 0.0021 & 0.0038 & 0.0048 & 0.0023 & 0.0023 \\
PathRAG           & 0.0009 & 0.0026 & 0.0030 & 0.0048 & 0.0021 & 0.0021 \\
CypherRAG         & 0.0004 & 0.0023 & 0.0028 & 0.0045 & 0.0014 & 0.0016 \\
\midrule
Multimodal Context & 0.0005 & 0.0022 & 0.0048 & 0.0046 & 0.0012 & 0.0018 \\
Proteus Context   & 0.0303 & 0.0393 & 0.0400 & 0.0363 & 0.0358 & 0.0342 \\
\midrule
\multicolumn{7}{c}{\textbf{execution time (s)}} \\
\midrule
ContextRAG        & 7.49s & 34.36s & 62.53s & 48.86s & 24.91s & 24.33s \\
VectorRAG         & 14.04s & 20.52s & 31.93s & 36.16s & 34.05s & 24.42s \\
PathRAG           & 51.62s & 40.54s & 24.29s & 45.92s & 76.41s & \underline{54.64s} \\
CypherRAG         & 19.35s & 51.99s & 52.30s & 74.94s & 42.60s & 39.42s \\
\midrule
Multimodal Context & 18.48s & 42.67s & 100.35s & 103.02s & 44.59s & 45.55s \\
Proteus Context   & 15.32s & 96.70s & 140.49s & 69.96s & 64.52s & 52.05s \\
\bottomrule
\end{tabularx}
\end{table}

ContextRAG consistently achieves the highest overall accuracy (0.91) across all task types, with a low cost per task (\$0.004) and a fast execution time (24s). VectorRAG and PathRAG have similar average accuracies (0.82-0.83), but their performance varies across tasks. VectorRAG is more accurate on graph query tasks, while PathRAG achieves better accuracy on graph summarization, knowledge inference, and path exploration. PathRAG achieves this higher score with a path-focused algorithm at the cost of longer execution time. 

CypherRAG performs well with an overall average score of 0.86, peaking in graph summarization (1.00) and single-graph queries (0.88). Among the baselines, the multimodal context is in the lower end, with an average accuracy score of 0.83. However, it remains a cost-effective alternative that performs well on high-level tasks such as summarization (0.95) and knowledge inference (0.91), though it struggles with precision-intensive tasks such as path exploration (0.80) and graph queries (0.79). In contrast, the Proteus Context achieved a high accuracy score of 0.88, but at a significantly higher cost (\$0.034 per task) and a longer average runtime of 52 seconds. A comprehensive breakdown of these results is available in the Appendix~\ref{app:complete results}.

\section{Discussions}
\label{sec:Discussions}

In this section, we analyze how variations in P\&ID representation, querying tools, and LLM selection influence the observed results and discuss the implications for system design, efficiency, and accuracy.

\subsection{Role of representation in performance and cost}
\label{subsec: Role of representation in performance and cost}

Our findings indicate that P\&ID representation provides modest improvements in accuracy for frontier models but yields substantial cost savings. Frontier models, such as GPT-5, possess robust reasoning capabilities for complex patterns~\citep{du2025ockbench, ke2025survey}, allowing them to handle dense information sources, such as Proteus XML or complete-level graphs. However, this depth comes at a significant computational premium. For instance, processing a Proteus context requires 150K input tokens (around \$0.175 per task in GPT-5), whereas a complete graph uses 66K tokens (\$0.113) and a process graph utilizes 32K tokens (\$0.055). At the highest level of abstraction, a conceptual graph requires only 7K tokens (\$0.027). Beyond the cost reduction, there is also a modest improvement (by 5\%) in response accuracy when using a conceptual graph than the Proteus context. 

By contrast, for smaller, more economical models such as GPT-4o mini, the input representation plays a much more decisive role in accuracy. Semantically dense and curated graph contexts help these models by removing noise and highlighting critical relationships. Proteus files often contain machine-oriented artifacts, such as URIs and structural metadata, which introduce "semantic noise" that can distort meaningful information transfer~\citep{noruzisemantic, hockett1952approach}. Smaller models frequently struggle to filter out this noise, whereas flagship models can navigate it successfully. This suggests a diverging design requirement: while larger models benefit from abstraction primarily for cost-efficiency, smaller models require high-level abstraction simply to maintain operational accuracy.

The limitations of multimodal representation underscore a trade-off between cost and precision. Although raw image inputs incur minimal token costs, they suffer from substantial performance penalties for specific graph-based queries. A common failure mode involves misreading textual data from the images; for example, in Figure~\ref{fig: Case PSV}~(iii), the model failed to correctly identify PSV set pressure. The internal pre-processing and downscaling inherent in LLM vision encoders likely contributed to these errors. However, there is no publicly available information on how this internal image processing is performed; hence, it is difficult to improve it. Furthermore, it is worth noting that the P\&ID image we use in this study is a relatively simple, high-resolution, single-page document (5,485 by 3,186 pixels at 330 DPI, 1.2MB). In contrast, typical industrial P\&IDs are far denser, with more objects and details per page, resulting in smaller font and symbol sizes. This may further reduce accuracy when represented as an image. Thus, while multimodal inputs remain satisfactory for general image summarization for high-level information, they are currently insufficient for tasks requiring high-fidelity data extraction. Overall, these findings underscore that careful representation design is essential for optimizing both cost and accuracy, particularly when deploying smaller models or large-scale multi-agent systems, especially in process engineering, where incorrect information can have severe consequences. 

Overall, these findings highlight that careful representation design can lead to meaningful improvements in both cost efficiency and task accuracy, particularly when deploying models or operating large-scale multi-agent systems.

\subsection{Strengths and limitations of tooling approaches}
\label{subsec: Strengths and limitations of tooling approaches}

The evaluation of ContextRAG demonstrates the effectiveness of graph abstraction for efficient knowledge retrieval, providing a robust balance of accuracy, cost, and execution speed. By pre-processing the graph to be semantically dense and filtering out irrelevant metadata (semantic noise), ContextRAG minimizes the number of processed tokens. However, its primary limitation emerges when flowsheets span multiple pages; because the entire graph must be passed to the LLM for each inference, costs scale linearly with graph size. To mitigate this, we implemented a "topology mode" that passes only the lightweight structural framework without data attributes. While this reduces overhead, extracting specific parameters still requires specialized tooling, necessitating more granular methods such as VectorRAG, PathRAG, and CypherRAG.

A comparison between VectorRAG and PathRAG reveals distinct operational strengths despite their shared vector-based foundation. PathRAG employs iterative traversal to improve its ability to locate specific technical nodes, but this precision comes at the cost of higher latency (averaging 54.6s per query). Conversely, VectorRAG leverages precomputed embeddings to cover a broader semantic space more rapidly and at a lower per-query cost. These methodologies represent an evolving frontier in graph-based retrieval, with significant potential for refinement in future research, alongside emerging research on graph traversal~\citep{edge2024local, sun2023think, ma2024think, he2024give}.

CypherRAG demonstrates that schema-aware querying can achieve reasonable accuracy, though its performance is heavily dependent on the underlying model. In our evaluation, only frontier models such as GPT-5 and GPT-5-mini achieve average accuracies above 0.75. In the experiment, to test the model's capability, we limit the query generation to a single attempt without iterative refinement. Because LLMs lack innate knowledge of the global graph's connectivity during initial query generation, query errors are frequent. A standard mitigation strategy integrates database-aware verification and syntactic feedback loops, enabling the LLM (or a human-in-the-loop) to refine the query until it is valid. This feedback-and-iteration approach has also been highlighted in prior studies~\citep{gusarov2025multi, yang2025askgraph, gupta2025pidqa}. However, as P\&ID complexity increases due to additional components and process variations, the graph size grows substantially, even if the schema remains relatively similar. This structural expansion complicates Cypher formulation; while the LLM may recognize all possible connections, it often struggles to identify the specific path required, necessitating multiple reasoning cycles to produce a functional query. Meanwhile, in this study, we used a standard Neo4j implementation; further customizing the Cypher generation for the pyDEXPI graph could significantly improve the robustness of LLM query generation.

Ultimately, while this evaluation focuses on a single-page, simple P\&ID, vector indices and schema-based methods (VectorRAG, PathRAG, and CypherRAG) will become increasingly important as diagrams grow in scale. These approaches possess the unique potential to offload query processing from the LLM to the database layer, which is essential for maintaining performance in large-scale industrial applications.

\subsection{Considerations for LLM selection, common failure, and deployment}
\label{subsec: Considerations for LLM selection, deployment, and common failure modes}
In Section~\ref{subsec:LLM model evaluation and selection}, we showed that LLM performance reflects a clear trade-off between model size, accuracy, and computational cost across task types. Here, we highlight practical insights for deploying LLMs in agentic systems for P\&ID tasks, as well as common failure modes observed during deployment.

LLM tasks for P\&ID systems span a wide spectrum, ranging from highly precise information queries that require accuracy to higher-level summarization tasks that demand broader graph-level understanding. Larger models are generally advantageous due to their longer context windows and greater model complexity, but at the cost of computational power. Meanwhile, smaller online models remain highly capable but are constrained by shorter context lengths, requiring careful context management. In the flowsheet setting, this implies maintaining dense, noise-free graph representations to maximize the amount of useful context. Offline models, although limited to smaller sizes in our experiments ($≤$20B parameters), can also be viable when combined with effective retrieval mechanisms. In particular, structured representations such as VectorRAG and PathRAG can compensate for limited model size by improving retrieval accuracy. In these tool-usage cases, more of the decision-making is handled by deterministic algorithms rather than the LLM, reducing the need for more complex models.

Examining agent states in our evaluation bench reveals several recurring failure modes in LLM-based workflows:
\begin{itemize}
\item The model is too small: Smaller models may struggle to interpret complex GraphML or Proteus XML structures, leading to incomplete or incorrect outputs. Similar effects are observed in multimodal models.
\item Too much semantic noise: Machine-oriented artifacts (e.g., URIs, raw identifiers, structural metadata) introduce semantic noise that smaller models in particular struggle to filter.
\item Context is too long: While modern models support long context windows and caching methods, using the entire context window does not necessarily improve accuracy and often increases token cost and inference time.
\item Infinite tools iteration: Tools can improve performance, but poorly constrained agents may invoke unproductive tool-call loops that exhaust context, inflate token costs, and significantly increase execution time. Prompt strategy constraints and tool-call limiters are required to prevent this failure mode.
\item Underspecified response formats: Without explicit output constraints, agents may produce verbose and unfocused answers, commonly known as LLMs yapping~\citep{borisov2026chatbot}. Specifying the preferred response format via a system prompt and enforcing structured outputs via schemas mitigate this behavior.
\item Impractical time: Offline models, such as those deployed via Ollama, frequently fail to complete complex tasks within practical time limits when run on a local desktop machine.
\item Inefficient prompt reuse and caching: Repeatedly passing large static artifacts (e.g., P\&ID diagrams or lengthy system prompts) in every inference incurs unnecessary token cost and latency. Modern frontier LLM platforms (e.g., OpenAI GPT, Anthropic Claude, Google Gemini) support prompt caching or persistent context mechanisms - some of which are automatic, others that require explicit configuration. Leveraging these features amortizes token usage across multi-step workflows, significantly reducing long-term costs.
\end{itemize}

During deployment, managing slow execution times and achieving reliable end-to-end accuracy often requires substantial compute resources. To avoid a sluggish user experience, it is critical to provide continuous feedback on task progress, including which tools are invoked, the agent's current state, and the latest execution status. Hence, in our ChatP\&ID, we use the streaming token method and provide information on the latest workflow status.

To further optimize execution time and accuracy, we also suggest that multi-agent architectures may offer a promising deployment strategy. In such configurations, a more capable online model can act as a supervisory coordinator, overseeing multiple tool-augmented agents executing queries in parallel~\citep{rupprechtMultiagentSystemsChemical2025}. Their outputs can then be compared, verified, or aggregated by the supervisor agent to improve both accuracy and robustness.

\section{Conclusion}
\label{sec:Conclusion}
We introduced ChatP\&ID, a framework that enables intuitive interaction with smart P\&IDs using GraphRAG and knowledge graphs. In this framework, DEXPI P\&IDs are encoded as a knowledge graph, and state-of-the-art GraphRAG methods are implemented to allow LLMs to query information directly from the P\&ID graph database. We evaluated the framework across diverse query tasks, graph abstractions, GraphRAG methods, and LLM selections, benchmarking performance against baselines that provide raw P\&ID images or DEXPI smart P\&ID files directly to the LLM. This work demonstrates the potential of GraphRAG for engineering diagrams by enabling grounded natural-language interaction with structured P\&IDs.

Our results demonstrate that graph-based representation improves LLM accuracy while significantly reducing token costs. Among the GraphRAG methods tested (ContextRAG, VectorRAG, PathRAG, and CypherRAG), ContextRAG achieved the highest accuracy, lowest cost per task, and fastest execution time. For offline models, vector-based approaches such as VectorRAG and PathRAG further improve accuracy by shifting the workload to vector-similarity algorithms. Additionally, we found that using LLMs as judges yields more reliable scores than traditional semantic similarity metrics, which measure semantic distance but do not always capture logical correctness.

During development and testing, we identified opportunities to further enhance framework accuracy, such as implementing a multi-agent system in which multiple tools operate concurrently, overseen by a supervisor agent that makes decisions based on worker-agent outputs~\citep{rupprechtMultiagentSystemsChemical2025}. While CypherRAG already performs well, its effectiveness could be further improved by designing a specialized Cypher query generator tailored for pyDEXPI graphs. Future work will expand these methods into broader tools, including flowsheet modification, and extend their application to help process engineers with daily tasks such as HAZOPs.

\section*{Acknowledgement}
\label{sec:Acknowledgement}
We gratefully acknowledge the support from the Indonesia Endowment Fund for Education Agency (LPDP; AAA, 202401220100409).

\bibliography{references}  

@article{borisov2026chatbot,
  title={Do Chatbot LLMs Talk Too Much? The YapBench Benchmark},
  author={Borisov, Vadim and Gr{\"o}ger, Michael and Mikhael, Mina and Schreiber, Richard H},
  journal={arXiv preprint arXiv:2601.00624},
  year={2026}
}

@article{wei2022chain,
  title={Chain-of-thought prompting elicits reasoning in large language models},
  author={Wei, Jason and Wang, Xuezhi and Schuurmans, Dale and Bosma, Maarten and Xia, Fei and Chi, Ed and Le, Quoc V and Zhou, Denny and others},
  journal={Advances in neural information processing systems},
  volume={35},
  pages={24824--24837},
  year={2022}
}

@inproceedings{baken2020linked,
  title={Linked data for smart homes: Comparing RDF and labeled property graphs},
  author={Baken, Nico},
  booktitle={LDAC2020—8th Linked Data in Architecture and Construction Workshop},
  pages={23--36},
  year={2020}
}

@article{di2023lpg,
  title={Lpg-based knowledge graphs: A survey, a proposal and current trends},
  author={Di Pierro, Davide and Ferilli, Stefano and Redavid, Domenico},
  journal={Information},
  volume={14},
  number={3},
  pages={154},
  year={2023},
  publisher={MDPI}
}

@article{tan2024judgebench,
  title={Judgebench: A benchmark for evaluating llm-based judges},
  author={Tan, Sijun and Zhuang, Siyuan and Montgomery, Kyle and Tang, William Y and Cuadron, Alejandro and Wang, Chenguang and Popa, Raluca Ada and Stoica, Ion},
  journal={arXiv preprint arXiv:2410.12784},
  year={2024}
}

@article{zheng2023judging,
  title={Judging llm-as-a-judge with mt-bench and chatbot arena},
  author={Zheng, Lianmin and Chiang, Wei-Lin and Sheng, Ying and Zhuang, Siyuan and Wu, Zhanghao and Zhuang, Yonghao and Lin, Zi and Li, Zhuohan and Li, Dacheng and Xing, Eric and others},
  journal={Advances in neural information processing systems},
  volume={36},
  pages={46595--46623},
  year={2023}
}

@article{gu2024survey,
  title={A survey on llm-as-a-judge},
  author={Gu, Jiawei and Jiang, Xuhui and Shi, Zhichao and Tan, Hexiang and Zhai, Xuehao and Xu, Chengjin and Li, Wei and Shen, Yinghan and Ma, Shengjie and Liu, Honghao and others},
  journal={The Innovation},
  year={2024},
  publisher={Elsevier}
}

@misc{langchain_neo4j_cypher_docs,
  title        = {Neo4j Cypher Integration — LangChain Documentation},
  author       = {{LangChain Contributors}},
  year         = {2025},
  url          = {https://docs.langchain.com/oss/python/integrations/graphs/neo4j_cypher/},
  note         = {Accessed: 2025-02-03},
  organization = {LangChain},
  description  = {Documentation page describing the Neo4j Cypher integration for Python in LangChain},
}

@article{shi2025deep,
  title={Deep semantic graph learning via llm based node enhancement},
  author={Shi, Chuanqi and Tao, Yiyi and Zhang, Hang and Wang, Lun and Du, Shaoshuai and Shen, Yixian and Shen, Yanxin},
  journal={arXiv preprint arXiv:2502.07982},
  year={2025}
}

@article{russell1995modern,
  title={A modern approach},
  author={Russell, Stuart and Norvig, Peter and Intelligence, Artificial},
  journal={Artificial Intelligence. Prentice-Hall, Egnlewood Cliffs},
  volume={25},
  number={27},
  pages={79--80},
  year={1995}
}

@misc{langgraph2024,
  author = {LangChain},
  title = {LangGraph: Building resilient language agents as graphs},
  year = {2024},
  url = {https://github.com/langchain-ai/langgraph},
  note = {Accessed: 2026-02-18}
}

@misc{langsmith2024,
  author = {LangChain},
  title = {LangSmith: A platform for building, testing, and monitoring LLM applications},
  year = {2024},
  url = {https://docs.langchain.com/langsmith/home},
  note = {Accessed: 2026-02-18}
}

@inproceedings{nam2024using,
  title={Using an llm to help with code understanding},
  author={Nam, Daye and Macvean, Andrew and Hellendoorn, Vincent and Vasilescu, Bogdan and Myers, Brad},
  booktitle={Proceedings of the IEEE/ACM 46th International Conference on Software Engineering},
  pages={1--13},
  year={2024}
}

@inproceedings{derose2024can,
  title={Can LLMs help with XML?},
  author={DeRose, Steven J},
  booktitle={Balisage: The Markup Conference},
  year={2024}
}

@article{butler2025massive,
  title={The Massive Legal Embedding Benchmark (MLEB)},
  author={Butler, Umar and Butler, Abdur-Rahman and Malec, Adrian Lucas},
  journal={arXiv preprint arXiv:2510.19365},
  year={2025}
}

@article{ponwitayarat2025sea,
  title={SEA-BED: Southeast Asia Embedding Benchmark},
  author={Ponwitayarat, Wuttikorn and Ng, Raymond and Montalan, Jann Railey and Aung, Thura and Ngui, Jian Gang and Susanto, Yosephine and Tjhi, William and Tasawong, Panuthep and Cambria, Erik and Chuangsuwanich, Ekapol and others},
  journal={arXiv preprint arXiv:2508.12243},
  year={2025}
}

@article{goel2025sage,
  title={SAGE: A Realistic Benchmark for Semantic Understanding},
  author={Goel, Samarth and Lee, Reagan J and Ramchandran, Kannan},
  journal={arXiv preprint arXiv:2509.21310},
  year={2025}
}

@book{barrasa2023building,
  title={Building knowledge graphs},
  author={Barrasa, Jes{\'u}s and Webber, Jim},
  year={2023},
  publisher={" O'Reilly Media, Inc."}
}

@incollection{pan2009resource,
  title={Resource description framework},
  author={Pan, Jeff Z},
  booktitle={Handbook on ontologies},
  pages={71--90},
  year={2009},
  publisher={Springer}
}

@article{schulze2025rule,
  title={Rule-based autocorrection of Piping and Instrumentation Diagrams (P\&IDs) on graphs},
  author={Schulze Balhorn, Lukas and Seijsener, Niels and Dao, Kevin and Kim, Minji and Goldstein, Dominik P and Driessen, Ge HM and Schweidtmann, Artur M},
  journal={arXiv e-prints},
  pages={arXiv--2502},
  year={2025}
}

@article{brown2020language,
  title={Language models are few-shot learners},
  author={Brown, Tom and Mann, Benjamin and Ryder, Nick and Subbiah, Melanie and Kaplan, Jared D and Dhariwal, Prafulla and Neelakantan, Arvind and Shyam, Pranav and Sastry, Girish and Askell, Amanda and others},
  journal={Advances in neural information processing systems},
  volume={33},
  pages={1877--1901},
  year={2020}
}

@inproceedings{yao2022react,
  title={React: Synergizing reasoning and acting in language models},
  author={Yao, Shunyu and Zhao, Jeffrey and Yu, Dian and Du, Nan and Shafran, Izhak and Narasimhan, Karthik R and Cao, Yuan},
  booktitle={The eleventh international conference on learning representations},
  year={2022}
}

@article{bran2023chemcrow,
  title={Chemcrow: Augmenting large-language models with chemistry tools},
  author={Bran, Andres M and Cox, Sam and Schilter, Oliver and Baldassari, Carlo and White, Andrew D and Schwaller, Philippe},
  journal={arXiv preprint arXiv:2304.05376},
  year={2023}
}

@article{theisen2025graph,
  title={Graph neural networks for soft sensors: Learning from process topology and operational data},
  author={Theisen, Maximilian F and Meesters, Gabrie MH and Schweidtmann, Artur M},
  journal={Computers \& Chemical Engineering},
  pages={109532},
  year={2025},
  publisher={Elsevier}
}

@article{balhorn2025graph,
  title={Graph-to-SFILES: Control structure prediction from process topologies using generative artificial intelligence},
  author={Balhorn, Lukas Schulze and Degens, Kevin and Schweidtmann, Artur M},
  journal={Computers \& Chemical Engineering},
  pages={109121},
  year={2025},
  publisher={Elsevier}
}

@article{li2020survey,
  title={A survey on deep learning for named entity recognition},
  author={Li, Jing and Sun, Aixin and Han, Jianglei and Li, Chenliang},
  journal={IEEE transactions on knowledge and data engineering},
  volume={34},
  number={1},
  pages={50--70},
  year={2020},
  publisher={IEEE}
}

@article{gowaikar2024agentic,
  title={An agentic approach to automatic creation of p\&id diagrams from natural language descriptions},
  author={Gowaikar, Shreeyash and Iyengar, Srinivasan and Segal, Sameer and Kalyanaraman, Shivkumar},
  journal={arXiv preprint arXiv:2412.12898},
  year={2024}
}

@inproceedings{mukharror2025use,
  title={On the Use of Artificial Intelligent with Large Language Model as a Tool for Evaluating Quality of Hazard and Operability Study},
  author={Mukharror, Darmawan Ahmad and Susilo, Henry and Putra, Zulfan Adi},
  booktitle={2025 9th International Conference on Instrumentation, Control, and Automation (ICA)},
  pages={337--342},
  year={2025},
  organization={IEEE}
}

@article{gupta2025pidqa,
  title={PIDQA—Question Answering on Piping and Instrumentation Diagrams},
  author={Gupta, Mohit and Wei, Chialing and Czerniawski, Thomas and Eiris, Ricardo},
  journal={Machine Learning and Knowledge Extraction},
  volume={7},
  number={2},
  pages={39},
  year={2025},
  publisher={MDPI}
}

@article{medhane2025automated,
  title={Automated counting of piping and instrumentation diagram using artificial intelligence},
  author={Medhane, Sampat and Khalkar, Rohini and Dhumane, Amol and Mestry, Rutuj and Khan, Aquib and Agarwal, Sajal},
  journal={Journal of Integrated Science and Technology},
  volume={13},
  number={6},
  pages={1147--1147},
  year={2025}
}

@article{LEE2026107039,
    title = {Can large language models automate the HAZOP process without human intervention?},
    journal = {Safety Science},
    volume = {194},
    pages = {107039},
    year = {2026},
    issn = {0925-7535},
    doi = {https://doi.org/10.1016/j.ssci.2025.107039},
    url = {https://www.sciencedirect.com/science/article/pii/S0925753525002644},
    author = {Junseo Lee and Sunhwa Park and Sehyeon Oh and Byungchol Ma}
}

@book{latora2017complex,
  title={Complex networks: principles, methods and applications},
  author={Latora, Vito and Nicosia, Vincenzo and Russo, Giovanni},
  year={2017},
  publisher={Cambridge University Press}
}

@inproceedings{yang2025askgraph,
  title={AskGraph: A Dependency-Aware Code Assistant Powered by Code Graphs and LLM-Generated Cypher Queries},
  author={Yang, Nan and Reynolds, Joseph and Prast, Laurens and Corvino, Rosilde},
  booktitle={2025 IEEE International Conference on Software Maintenance and Evolution (ICSME)},
  pages={644--655},
  year={2025},
  organization={IEEE}
}

@article{noruzisemantic,
  title={Semantic Noise in Information Representation and Retrieval},
  author={Noruzi, Alireza},
  journal={informology},
  year={2024}
}

@article{hockett1952approach,
  title={An approach to the quantification of semantic noise},
  author={Hockett, Charles F},
  journal={Philosophy of science},
  volume={19},
  number={4},
  pages={257--260},
  year={1952},
  publisher={Cambridge University Press}
}

@article{ke2025survey,
  title={A survey of frontiers in llm reasoning: Inference scaling, learning to reason, and agentic systems},
  author={Ke, Zixuan and Jiao, Fangkai and Ming, Yifei and Nguyen, Xuan-Phi and Xu, Austin and Long, Do Xuan and Li, Minzhi and Qin, Chengwei and Wang, Peifeng and Savarese, Silvio and others},
  journal={arXiv preprint arXiv:2504.09037},
  year={2025}
}

@article{du2025ockbench,
  title={OckBench: Measuring the Efficiency of LLM Reasoning},
  author={Du, Zheng and Kang, Hao and Han, Song and Krishna, Tushar and Zhu, Ligeng},
  journal={arXiv preprint arXiv:2511.05722},
  year={2025}
}

@article{schulze2024empirical,
  title={Empirical assessment of ChatGPT’s answering capabilities in natural science and engineering},
  author={Schulze Balhorn, Lukas and Weber, Jana M and Buijsman, Stefan and Hildebrandt, Julian R and Ziefle, Martina and Schweidtmann, Artur M},
  journal={Scientific Reports},
  volume={14},
  number={1},
  pages={4998},
  year={2024},
  publisher={Nature Publishing Group UK London}
}

@book{ghallab2004automated,
  title={Automated Planning: theory and practice},
  author={Ghallab, Malik and Nau, Dana and Traverso, Paolo},
  year={2004},
  publisher={Elsevier}
}

@article{edge2024local,
  title={From local to global: A graph rag approach to query-focused summarization},
  author={Edge, Darren and Trinh, Ha and Cheng, Newman and Bradley, Joshua and Chao, Alex and Mody, Apurva and Truitt, Steven and Metropolitansky, Dasha and Ness, Robert Osazuwa and Larson, Jonathan},
  journal={arXiv preprint arXiv:2404.16130},
  year={2024}
}

@article{liu2024lost,
  title={Lost in the middle: How language models use long contexts},
  author={Liu, Nelson F and Lin, Kevin and Hewitt, John and Paranjape, Ashwin and Bevilacqua, Michele and Petroni, Fabio and Liang, Percy},
  journal={Transactions of the Association for Computational Linguistics},
  volume={12},
  pages={157--173},
  year={2024}
}

@article{kuratov2024search,
  title={In search of needles in a 11m haystack: Recurrent memory finds what llms miss},
  author={Kuratov, Yuri and Bulatov, Aydar and Anokhin, Petr and Sorokin, Dmitry and Sorokin, Artyom and Burtsev, Mikhail},
  journal={arXiv preprint arXiv:2402.10790},
  year={2024}
}

@inproceedings{hasan2025engineering,
  title={Engineering RAG Systems for Real-World Applications: Design, Development, and Evaluation},
  author={Hasan, Md Toufique and Waseem, Muhammad and Kemell, Kai-Kristian and Khan, Ayman Asad and Saari, Mika and Abrahamsson, Pekka},
  booktitle={Euromicro Conference on Software Engineering and Advanced Applications},
  pages={143--158},
  year={2025},
  organization={Springer}
}

@article{gao2023retrieval,
  title={Retrieval-augmented generation for large language models: A survey},
  author={Gao, Yunfan and Xiong, Yun and Gao, Xinyu and Jia, Kangxiang and Pan, Jinliu and Bi, Yuxi and Dai, Yixin and Sun, Jiawei and Wang, Haofen and Wang, Haofen},
  journal={arXiv preprint arXiv:2312.10997},
  volume={2},
  number={1},
  year={2023}
}

@article{dang2025survey,
  title={Survey and analysis of hallucinations in large language models: attribution to prompting strategies or model behavior},
  author={Dang, Hoang Anh and Tran, Vu and Nguyen, Le-Minh},
  journal={Frontiers in Artificial Intelligence},
  volume={8},
  pages={1622292},
  year={2025},
  publisher={Frontiers}
}

@article{lewis2020retrieval,
  title={Retrieval-augmented generation for knowledge-intensive nlp tasks},
  author={Lewis, Patrick and Perez, Ethan and Piktus, Aleksandra and Petroni, Fabio and Karpukhin, Vladimir and Goyal, Naman and K{\"u}ttler, Heinrich and Lewis, Mike and Yih, Wen-tau and Rockt{\"a}schel, Tim and others},
  journal={Advances in neural information processing systems},
  volume={33},
  pages={9459--9474},
  year={2020}
}

@article{vaswani2017attention,
  title={Attention is all you need},
  author={Vaswani, Ashish and Shazeer, Noam and Parmar, Niki and Uszkoreit, Jakob and Jones, Llion and Gomez, Aidan N and Kaiser, {\L}ukasz and Polosukhin, Illia},
  journal={Advances in neural information processing systems},
  volume={30},
  year={2017}
}

@article{radford2018improving,
  title={Improving language understanding with unsupervised learning},
  author={Radford, Alec},
  journal={OpenAI Res},
  year={2018}
}

@article{szlobodnyik5669662feedback,
  title={Feedback-Guided Prompt Injection Defense in Retrieval-Augmented Text-to-Cypher Generation},
  author={Szlobodnyik, Gergely},
  journal={Available at SSRN 5669662},
  year={2025}
}

@article{gusarov2025multi,
  title={Multi-Agent GraphRAG: A Text-to-Cypher Framework for Labeled Property Graphs},
  author={Gusarov, Anton and Volkova, Anastasia and Khrulkov, Valentin and Kuznetsov, Andrey and Maslov, Evgenii and Oseledets, Ivan},
  journal={arXiv preprint arXiv:2511.08274},
  year={2025}
}

@article{sun2023think,
  title={Think-on-graph: Deep and responsible reasoning of large language model on knowledge graph},
  author={Sun, Jiashuo and Xu, Chengjin and Tang, Lumingyuan and Wang, Saizhuo and Lin, Chen and Gong, Yeyun and Ni, Lionel M and Shum, Heung-Yeung and Guo, Jian},
  journal={arXiv preprint arXiv:2307.07697},
  year={2023}
}

@article{ma2024think,
  title={Think-on-graph 2.0: Deep and interpretable large language model reasoning with knowledge graph-guided retrieval},
  author={Ma, Shengjie and Xu, Chengjin and Jiang, Xuhui and Li, Muzhi and Qu, Huaren and Guo, Jian},
  journal={arXiv e-prints},
  pages={arXiv--2407},
  year={2024}
}

@article{he2024give,
  title={Give: Structured reasoning of large language models with knowledge graph inspired veracity extrapolation},
  author={He, Jiashu and Ma, Mingyu Derek and Fan, Jinxuan and Roth, Dan and Wang, Wei and Ribeiro, Alejandro},
  journal={arXiv preprint arXiv:2410.08475},
  year={2024}
}

@article{alimin2025talking,
  title={Talking like piping and instrumentation diagrams (p\&ids)},
  author={Alimin, Achmad Anggawirya and Goldstein, Dominik P and Balhorn, Lukas Schulze and Schweidtmann, Artur M},
  journal={arXiv preprint arXiv:2502.18928},
  year={2025}
}

@misc{rupprechtMultiagentSystemsChemical2025,
  title = {Multi-Agent Systems for Chemical Engineering: {{A}} Review and Perspective},
  shorttitle = {Multi-Agent Systems for Chemical Engineering},
  author = {Rupprecht, Sophia and Gao, Qinghe and Karia, Tanuj and Schweidtmann, Artur M.},
  year = {2025},
  month = aug,
  number = {arXiv:2508.07880},
  eprint = {2508.07880},
  primaryclass = {cs},
  publisher = {arXiv},
  doi = {10.48550/arXiv.2508.07880},
  urldate = {2025-08-18},
  archiveprefix = {arXiv},
  langid = {english}
}

@article{morbach2007ontocape,
  title={OntoCAPE—A large-scale ontology for chemical process engineering},
  author={Morbach, Jan and Yang, Aidong and Marquardt, Wolfgang},
  journal={Engineering applications of artificial intelligence},
  volume={20},
  number={2},
  pages={147--161},
  year={2007},
  publisher={Elsevier}
}

@article{schweidtmann2024generative,
  title={Generative artificial intelligence in chemical engineering},
  author={Schweidtmann, Artur M},
  journal={Nature Chemical Engineering},
  volume={1},
  number={3},
  pages={193--193},
  year={2024},
  publisher={Nature Publishing Group US New York}
}

@article{theisen2023digitization,
  title={Digitization of chemical process flow diagrams using deep convolutional neural networks},
  author={Theisen, Maximilian F and Flores, Kenji Nishizaki and Schulze Balhorn, Lukas and Schweidtmann, Artur M},
  journal={Digital Chemical Engineering},
  volume={6},
  pages={100072},
  year={2023},
  publisher={Elsevier}
}

@book{toghraei2019piping,
  title={Piping and instrumentation diagram development},
  author={Toghraei, Moe},
  year={2019},
  publisher={John Wiley \& Sons}
}

@Manual{Theissen2021,
  title        = {{DEXPI P\&ID} specification},
  author       = {Theißen, Manfred and Wiedau, Michael},
  note         = {Version 1.3},
  organization = {DEXPI Initiative},
  year         = {2021},
  ranking      = {rank4},
}

@InProceedings{Goldstein2025,
  author    = {Goldstein, Dominik P. and Alimin, Achmad Anggawirya and Schulze Balhorn, Lukas and Schweidtmann, Artur M},
  booktitle = {Proceedings of the 35th European Symposium on Computer Aided Process Engineering (ESCAPE35)},
  title     = {pyDEXPI:{A} {Python} framework for piping and instrumentation diagrams using the {DEXPI} information model},
  year      = {2025},
  address   = {Ghent, Belgium},
  month     = {July 6--9},
}

@article{eibeck_j-park_2019,
    title = {J-{Park} {Simulator}: {An} ontology-based platform for cross-domain scenarios in process industry},
    volume = {131},
    issn = {00981354},
    shorttitle = {J-{Park} {Simulator}},
    url = {https://linkinghub.elsevier.com/retrieve/pii/S0098135419301589},
    doi = {10.1016/j.compchemeng.2019.106586},
    language = {en},
    urldate = {2025-01-28},
    journal = {Computers \& Chemical Engineering},
    author = {Eibeck, Andreas and Lim, Mei Qi and Kraft, Markus},
    month = dec,
    year = {2019},
    pages = {106586},
}

\newpage

\appendix
\label{appendix} 

\captionsetup[table]{name=App. Table, labelsep=colon, labelfont=bf}
\renewcommand{\thetable}{\arabic{table}}
\setcounter{table}{0}

\section{Generation of Local and Global Semantic Descriptions and Embedding}
\label{app:semantic_generation}
This appendix describes how semantic descriptions are generated for each node in the process flowsheet graph and embedded in the vector space. 
Two complementary semantic views are created:

\begin{itemize}
    \item \textbf{Global semantic:} the functional role of a node in the complete flowsheet.
    \item \textbf{Local semantic:} the immediate relational context of a node with its first-level neighbors.
\end{itemize}

Both descriptions are produced using an LLM that receives structured graph information extracted from Neo4j.

\subsection{Global Semantic Generation}

The global semantics describes the overall purpose of a component within the full process flowsheet. 
The LLM receives the node attributes together with the complete graph representation, enabling reasoning about system-wide function.

\paragraph{Global Semantic Prompt}

\begin{verbatim}
<instruction>
You are a process engineering assistant.
Describe the role and function of this node in the process flowsheet 
(global context), based on given information.
Focus on what the equipment/component does, using its name or tag 
(not IDs).
Be clear and concise.
</instruction>

<node>
Labels: {labels}
Properties: {properties_json}
This node is a component of a process flowsheet.
</node>

<flowsheet>
{full_graph_representation}
</flowsheet>
\end{verbatim}

\subsection{Local Semantic Generation}

The local semantics describes how a node interacts with its immediate neighbors, capturing direct process connectivity such as upstream and downstream relationships.

\paragraph{Local Semantic Prompt}

\begin{verbatim}
<instruction>
You are a process engineering assistant.
Describe the local context of this node in the process flowsheet, 
focusing on its immediate relationships.
Use the node's labels and properties, and explain how it connects 
to its neighbors.
Focus on what the equipment/component does, using its name or tag 
(not IDs).
Be clear and concise.
</instruction>

<node>
Central Node:
Labels: {node_labels}
Properties: {node_properties_json}
This node is a component of a process flowsheet.
</node>

<connections>
Incoming Connections:
{incoming_connections_json}

Outgoing Connections:
{outgoing_connections_json}
</connections>
\end{verbatim}

\subsection{Semantic Representation Output}

For each node $n$, the LLM returns:

\begin{itemize}
    \item $global\_description_n$ — functional role in full flowsheet
    \item $local\_description_n$ — interaction with neighbors
\end{itemize}

These are combined into a unified semantic representation:

\[
semantic_n = (global\_description_n,\; local\_description_n)
\]

This semantic tuple is subsequently encoded into a vector embedding, as described in Algorithm~\ref{alg:semantic_enrichment}.

\subsection{Example}

This is the example for Global and Local semantic of Tank T4750

\textbf{Global semantic:}

\begin{verbatim}
<data key="global_semantic">
  The component with the tag "T4750" is a Tank, designed as a vessel
  for fluid storage within the process flowsheet. Its main role is to
  receive and collect fluids, serving as an intermediate storage point
  before the fluid is distributed for further processing. The tank is
  cylindrical with a length of 4.0 meters. The tank receives fluid
  from upstream valves and distributes it downstream, functioning as a
  critical storage and transfer point in the overall material flow. It
  also integrates with process instrumentation that monitors
  temperature signals, indicating its importance in process control
  and safety operations.
  </data>
\end{verbatim}

\textbf{Global semantic embedding:}

\begin{verbatim}
<data key="global_semantic_embedding">[0.016520526,-0.014927468
-0.020687385,-0.0301709,-0.008441978,...,-0.012176747]
\end{verbatim}

\textbf{Local semantic:}

\begin{verbatim}
<data key="local_semantic">
  The central node in this process flowsheet is equipment "T4750",
  classified as a tank vessel used for fluid storage and distribution.

  Incoming connections include multiple globe valves supplying fluids
  (codes MNc and MNb) and a spring-loaded safety valve that relieves
  overpressure into the tank. Outgoing connections include a butterfly
  valve distributing stored fluid downstream and a process signal
  generating function (TT4750.03) providing instrumentation and
  monitoring. The tank contains two internal chambers with specified
  materials and pressure–temperature design limits.

  This configuration shows the tank acting as a receiving, storage,
  and controlled distribution node within the flowsheet.
  </data>
\end{verbatim}

\textbf{Local semantic embedding:}

\begin{verbatim}
<data key="local_semantic_embedding">[-3.1762512E-4,-0.006226967,
-0.02870905,-0.048853412,0.0532493,0.03661649,...,-0.01421397]
\end{verbatim}

\section{Dataset of Questions and Reference Answers for Graph Tasks}
\label{app:qa set}

\newcommand{\longHeader}{Task Type & Question & Reference Answer \\}
\begin{longtable}{p{2cm} p{4cm} p{9cm}}
\caption{Dataset of Questions and Reference Answers for Graph Tasks} \label{appendix:qa_dataset} \\

\toprule
\longHeader
\midrule
\endfirsthead
\multicolumn{3}{l}{\footnotesize \ldots App. Table~\ref{appendix:qa_dataset}, continued from the previous page.} \\
\midrule
\longHeader
\midrule
\endhead
\midrule
\multicolumn{3}{r}{\footnotesize Continues on the next page \ldots}
\endfoot
\bottomrule
\endlastfoot

Graph Query (Single)  & What is the cylinder length of tank T4750? & 4.0 m\newline\newline \\
Graph Query (Single) & What is the design heat transfer area of heat exchanger H1007? & 46.8 m$^2$\newline\newline \\
Graph Query (Single) & What is the design pressure of pump P4711? & 10.0 m (pressure head)\newline\newline \\
Graph Query (Single) & What is the design shaft power of pump P4711? & 60.0 kW\newline \\
Graph Query (Single) & What is the design volume flow rate of pump P4712? & 420.0 m$^3$/h\newline\newline \\
Graph Query (Single) & What is the nominal diameter of valve 66KL21? & DN 80\newline\newline \\
Graph Query (Single) & What is the set pressure of safety valve SV 104.01? & 6.0 bar\newline\newline \\
Graph Query (Single) & What is the upper design temperature of tank T4750? & 100.0 $^{\circ}$C\newline\newline \\
Graph Query (Multi) & 
List all valves in the P\&ID along with their specification. & 
1. Globe Valves \newline
Valve 1: \newline
- Type: Globe Valve (Operated) \newline
- Tag/Component: C1 \newline
- Nominal Diameter: DN 80 \newline
- Piping Class: 75HB13 \newline
- Fluid Code: MNb \newline
- Segment: S1 \newline
- Actuation: HV4750.01 (Fail Close) \newline \newline
Valve 2: \newline
- Type: Globe Valve (Operated) \newline
- Tag/Component: C1 \newline
- Nominal Diameter: DN 50 \newline
- Piping Class: 75HB13 \newline
- Fluid Code: MNc \newline
- Segment: S1 \newline
- Insulation: 80.0 mm thickness, Type Q \newline
- Actuation: PV4712.02 (Fail Close) \newline\newline
\\&&
Valve 3: \newline
- Type: Globe Valve (Operated) \newline
- Tag/Component: C1 \newline
- Nominal Diameter: DN 50 \newline
- Piping Class: 75HB13 \newline
- Fluid Code: QSb \newline
- Segment: S1 \newline
- Insulation: 80.0 mm thickness, Type Q \newline
- Actuation: TV4750.03 (Fail Open) \newline \newline
2. Butterfly Valves \newline
Valve 4: \newline
- Type: Butterfly Valve (Operated) \newline
- Component Name: 66KL21 \newline
- Tag/Component: C1 \newline
- Nominal Diameter: DN 80 \newline
- Piping Class: 75HB13 \newline
- Fluid Code: MNc \newline
- Segment: S1 \newline \newline
3. Check Valves \newline
Valve 5: \newline
- Type: Swing Check Valve \newline
- Component Name: 75SA21 \newline
- Tag/Component: C2 \newline
- Nominal Diameter: DN 80 \newline
- Piping Class: 75HB13 \newline
- Fluid Code: MNc \newline
- Segment: S1 \newline \newline
4. Ball Valves \newline
Valve 6: \newline
- Type: Ball Valve (Operated) \newline
- Component Name: 73KH12 \newline
- Tag/Component: C4 \newline
- Nominal Diameter: DN 50 \newline
- Piping Class: 73HG12 \newline
- Fluid Code: MNc \newline
- Line Number: 47124 \newline
- Segment: S3 \newline \newline
Valve 7: \newline
- Type: Ball Valve (Operated) \newline
- Component Name: 73KH12 \newline
- Tag/Component: C2 \newline
- Nominal Diameter: DN 25 \newline
- Piping Class: 75HB13 \newline
- Fluid Code: MNc \newline
- Line Number: 47126 \newline
- Segment: S2 \newline \newline
\\&&
Valve 8: \newline
- Type: Ball Valve (Operated) \newline
- Component Name: 73KH12 \newline
- Tag/Component: C5 \newline
- Nominal Diameter: DN 25 \newline
- Piping Class: 75HB13 \newline
- Fluid Code: MNc \newline
- Line Number: 47126 \newline
- Segment: S5 \newline \newline
Valve 9: \newline
- Type: Ball Valve (Operated) \newline
- Component Name: 73KH12 \newline
- Tag/Component: C7 \newline
- Nominal Diameter: DN 50 \newline
- Piping Class: 75HB13 \newline
- Fluid Code: MNc \newline
- Line Number: 47126 \newline
- Segment: S6 \newline \newline
Valve 10: \newline
- Type: Ball Valve (Operated) \newline
- Component Name: 73KH12 \newline
- Tag/Component: C10 \newline
- Nominal Diameter: DN 25 \newline
- Piping Class: 75HB13 \newline
- Fluid Code: MNc \newline
- Line Number: 47126 \newline
- Segment: S9 \newline \newline
5. Safety Valve \newline
Valve 11: \newline
- Position number: SV 104.01 \newline
- Type: SpringLoadedGlobeSafetyValve \newline
- Piping class: 73HG12 \newline
- Set pressure: 6.0 bar \newline
- Nominal diameter: DN 50 \newline
- Fluid code: MNc \newline
- Segment: S1. \newline\\
Graph Query (Multi) & 
List all pipe fitting in the P\&ID along with their specification. & 
1. Pipe Reducers \newline
- Type: PipeReducer \newline
- Component Number: C3 \newline
- Nominal Diameter: DN 50 \newline
- Piping Class Code: 73HG12 \newline
- Fluid Code: MNc \newline
- Line Number: 47124 \newline
- Segment Number: S2 \newline \newline
\\&&
2. Pipe Tees \newline
Tee 1: \newline
- Type: PipeTee \newline
- Component Number: C3 \newline
- Nominal Diameter: DN 25 \newline
- Piping Class Code: 75HB13 \newline
- Fluid Code: MNc \newline
- Line Number: 47126 \newline
- Segment Number: S3 \newline \newline
Tee 2: \newline
- Type: PipeTee \newline
- Component Number: C1 \newline
- Nominal Diameter: DN 25 \newline
- Piping Class Code: 75HB13 \newline
- Fluid Code: MNc \newline
- Segment Number: S1 \newline \newline
Tee 3: \newline
- Type: PipeTee \newline
- Component Number: C4 \newline
- Nominal Diameter: DN 25 \newline
- Piping Class Code: 75HB13 \newline
- Fluid Code: MNc \newline
- Line Number: 47126 \newline
- Segment Number: S4 \newline \newline
Tee 4: \newline
- Type: PipeTee \newline
- Component Number: C8 \newline
- Nominal Diameter: DN 50 \newline
- Piping Class Code: 75HB13 \newline
- Fluid Code: MNc \newline
- Line Number: 47126 \newline
- Segment Number: S6 \newline \newline
Tee 5: \newline
- Type: PipeTee \newline
- Component Number: C9 \newline
- Nominal Diameter: DN 25 \newline
- Piping Class Code: 75HB13 \newline
- Fluid Code: MNc \newline
- Line Number: 47126 \newline
- Segment Number: S8 \newline \newline
3. Blind Flanges \newline
Flange 1: \newline
- Type: BlindFlange \newline
- Component Number: C6 \newline
- Nominal Diameter: DN 25 \newline
- Piping Class Code: 75HB13 \newline
- Fluid Code: MNc \newline
- Line Number: 47126 \newline
- Segment Number: S5 \newline \newline
\\&&
Flange 2: \newline
- Type: BlindFlange \newline
- Component Number: C11 \newline
- Nominal Diameter: DN 25 \newline
- Piping Class Code: 75HB13 \newline
- Fluid Code: MNc \newline
- Line Number: 47126 \newline
- Segment Number: S9 \newline \\
Path Exploration & 
If heat exchanger H1007 needs maintenance, can flow still reach tank T4750? Describe the alternative path. & 
The operational strategy depends on the criticality of H1007. If raw material can be supplied at required temperature or no control is needed, H1007 could be bypassed. However, this requires a tie-in point along line MNb47123 which currently does not exist, or a connection via N3 of T4750. H1008 controls tank temperature, so using it as an alternative path is not recommended.\newline \\
Path Exploration & 
If you need to isolate tank T4750 from all upstream equipment, which valves would you need to close? & 
To fully isolate Tank T4750: \newline
- Shut the main DN80 feed globe valve (HV4750.01).\newline
- Close the DN50 globe valve on the recirculation line (PV4712.02).\newline \\
Path Exploration & 
Trace the flow path from tank T4750 to pump P4712. & 
Path: T4750 (Tank) $\rightarrow$ 66KL21 (Butterfly valve) $\rightarrow$ 75SA21 (Swing check valve) $\rightarrow$ Pipe reducer (DN80 to DN50) $\rightarrow$ 73KH12 (Ball valve) $\rightarrow$ P4712 (Reciprocating pump)\newline  \\
Path Exploration & 
Working backwards from tank T4750, what are the two possible inlet paths and their source equipment? & 
Path 1: T4750 $\leftarrow$ Globe Valve (HV4750.01) $\leftarrow$ H1007 (Plate Heat Exchanger) $\leftarrow$ P4711 (Centrifugal Pump) \newline
Path 2: T4750 $\leftarrow$ Globe Valve (TV4712.02) $\leftarrow$ H1008 (Tubular Heat Exchanger)\newline  \\
Path Exploration & 
What is the flow path from pump P4711 to tank T4750? & 
P4711 (Centrifugal Pump) $\rightarrow$ H1007 (Plate Heat Exchanger) $\rightarrow$ HV4750.01 (Globe Valve) $\rightarrow$ T4750 (Tank)\newline \\
Knowledge Inference & 
Analyze the flowsheet and give recommendations regarding process safety & 
Critical Safety Observations: \newline \newline
1. Safety valve set pressure vs vessel design pressure mismatch \newline
Current State: Safety valve SV104.01 is set at 6.0 bar and discharges to Tank T4750. However, the chambers of T4750 have upper design pressure limits of only 0.1 bar and 0.05 bar, indicating that the tank is essentially atmospheric or very low pressure. Under these conditions, a PSV set at 6 bar provides no effective protection to the tank. The tank would be subjected to severe overpressure and could fail catastrophically well before the safety valve opens. \newline
Recommendation: Each pressure-containing boundary must be protected by appropriately sized and correctly set overpressure protection devices. PSV set pressures shall not exceed the MAWP of the protected equipment. \newline \newline
\\&&
2. Missing pressure relief coverage for connected pressurized equipment \newline
Current State: Pumps P4711 and P4712, as well as heat exchangers H1007 and H1008, are specified with chamber design pressures reaching up to 60 bar. Despite this, pressure relief provisions for these high-pressure components are missing or unclear. \newline
Recommendation: Pressure relief analysis shall be updated to explicitly include operating conditions. Pressure relief devices shall be installed on all high-pressure equipment. \newline \newline
3. Process Fluid \newline
Current State: Multiple fluid codes are referenced (MNb, MNc, QSb) but no clear information is provided regarding their abbreviation. \newline
Recommendation: Provide detailed information on all process fluids and utility. Re-evaluate safety assessment based on confirmed fluids. \newline
4. Isolation \newline
Current State: Blind flanges are installed on several branch connections without clear isolation points. Major equipment items do not clearly show complete isolation points on all inlets and outlets. \newline
Recommendation: Isolation and line-breaking procedures should be formally reviewed. Positive isolation points should be provided for all major equipment. \newline \newline
5. Fail-Safe Action Analysis \newline
Current State: Valve fail actions are defined (PV4712.02 closed, HV4750.01 closed, TV4750.03 open). For TV4750.03, it is unclear if the fluid is heating or cooling, making fail-open potentially unsafe. \newline
Critical Recommendation: Comprehensive review of fail-to-safe-state philosophy for all critical valves.\newline \\
Knowledge Inference & 
List all control valves along with its control philosophy, why and what ifs fail. & 
Control Valves and Their Control Philosophies \newline \newline
1. HV4750.01 (Flow Control Valve) \newline
HV4750.01 is a manually operated globe valve used to regulate flow into Tank T4750. It is configured to fail closed. The fail-close action ensures a safe shutdown position by preventing uncontrolled flow into the tank in the event of actuator failure. If the valve fails, it closes and isolates flow to T4750, which may result in a low-level condition in the tank. \newline \newline
2. PV4712.02 (Pressure Control Valve) \newline
PV4712.02 is a globe valve used for pressure regulation within the downstream piping system and is controlled via a pressure control loop (PICSA 4712.02). It is also configured to fail closed. This fail-safe position prevents the process stream from returning to Tank T4750 if actuator power is lost. If the valve fails, flow through the return line is stopped, which can increase downstream pressure. \newline \newline
\\&&
3. TV4750.03 (Temperature Control Valve) \newline
TV4750.03 is a globe valve that regulates the flow of heating or cooling medium to heat exchanger H1008. It is configured to fail open. The fail-open design ensures continuous flow of the utility medium in case of actuator failure. If the valve fails, the medium flows at maximum capacity, which could lead to overcooling or overheating depending on the service.\newline \\
Knowledge Inference & 
What would be the effect if the heating fluid temperature in Heat Exchanger H1007 increased from the design temperature to 120$^{\circ}$C? & 
Context and Analysis Framework: \newline
H1007 is a Plate Heat Exchanger (Design: 313 kW, 46.8 m$^2$). \newline\newline 
Expected Effects: \newline
1. Heat Transfer Performance: Increased $\Delta$T would enhance heat transfer rate, potentially exceeding the design heat flow rate of 313 kW. Process fluid outlet temperature would increase. \newline\newline 
2. Equipment Safety: Chamber design temperature limits are 100$^{\circ}$C. Heating to 120$^{\circ}$C would exceed design temperature by 20$^{\circ}$C, risking thermal stress and equipment damage. \newline\newline 
3. Process Impact: Downstream equipment (Tank T4750) and processes would be affected. The temperature control loop (TV4750.03) would need to compensate. \newline\newline 
4. System Response: Control valve TV4750.03 might need to throttle more to maintain desired process temperature.\newline  \\
Graph Summarization & 
Based on the P\&ID, make a narrative to describe the process flow from the inlet to the final outlet, identifying all major equipment, intermediate streams, and control points in sequence. & 
Process Flow Description from Inlet to Outlet \newline \newline
1. Inlet and Feed Pressurization \newline
- The process fluid enters the system through the off-page inlet connector. From this inlet, the stream is routed to centrifugal feed pump P4711. This pump raises the fluid pressure to the level required for downstream operations. \newline \newline
2. First Heat Exchange \newline
- The discharge from P4711 is directed to plate heat exchanger H1007, where the fluid undergoes the initial heat transfer duty. \newline
- Downstream of H1007, flow passes through an actuated globe valve HS4750.01, which regulates the flow rate entering Tank T4750. \newline \newline
3. Tank \newline
- The controlled process stream enters Tank T4750. \newline
- The tank is fitted with a temperature transmitter and controller, TICSA4750.03, which continuously measures the vessel temperature and provides feedback to the temperature control loop. \newline \newline 
\\&&
4. Tank Outlet Isolation and Conditioning \newline
- Fluid exits the bottom of T4750 through a butterfly valve (66KL21). A swing check valve (75SA21) downstream prevents reverse flow. \newline
- The line size is then reduced from DN80 to DN50, and an operated ball valve (73KH12) provides additional isolation upstream of the downstream pump. \newline \newline
5. Secondary Pumping \newline
- The stream then continues to the reciprocating pump P4712, which delivers the pressurized stream into a downstream of the system. \newline \newline
6. Pump Discharge Protection and Distribution \newline
- Immediately downstream of P4712, the discharge enters a tee arrangement. \newline
- One branch contains a spring-loaded safety relief valve SV104.01, set at 6.0 bar, to protect the pump discharge piping from overpressure. Any relieved fluid is returned to Tank T4750. \newline
- The main discharge continues through additional tees, for pressure measurement and downstream process. \newline \newline
7. Instrumentation and Test Point \newline
- A branch from the post-pump tee passes through a ball valve to a blind flange equipped with pressure transmitter PI4712.01, providing local pressure measurement. \newline
- Further downstream, another tee divides the flow between the main process line and a controlled heat exchanger loop. \newline \newline
8. Tubular Heat Exchanger Loop and Tank Return \newline
- One branch from this tee supplies tubular heat exchanger H1008, which is part of the tank temperature control system. \newline
- The tank temperature controller TICSA 4750.03 modulates control valve TV4750.03 to regulate the flow of heating or cooling medium through H1008. \newline
- The process outlet from H1008 may be routed back to Tank T4750 via a controlled return line fitted with globe valve PV4712.02. \newline
- This recirculation loop enables control of the tank temperature. \newline \newline
9. Main Outlet Line \newline
- The main run from the tee bypasses H1008 and continues toward the system outlet. \newline
- Near the outlet, a small branch fitted with a ball valve and blind flange houses pressure transmitter and controller PICSA 4712.02. This instrument provides the pressure signal used to regulate the tank return line MNc47127 via control valve PV4712.02. \newline
- The main line terminates at the off-page outlet connector. \newline
\\&&
\textbf{Control and Instrumentation Summary} \newline
\textit{Flow Routing Control:} The actuated globe valve HS4750.01, located between H1007 and T4750, controls flow into the tank. The valve is operated from the central control system and is designed to fail closed. \newline\newline
\textit{Pressure Control:} PICSA 4712.02 measures pressure in the outlet line and generates the control signal for the tank return line. This controller modulates globe valve PV4712.02 in the H1008 return line, adjusting recirculation back to Tank T4750 to maintain the required line pressure. PV4712.02 is configured to fail closed. \newline\newline
\textit{Temperature Control:} TICSA 4750.03 measures the temperature in Tank T4750 and controls the temperature via the tank return loop. The controller actuates valve TV4750.03 to regulate the heating or cooling medium flow through H1008, maintaining the target tank temperature. TV4750.03 is designed to fail open. \newline\\

\bottomrule
\end{longtable}

\section{Complete Results}
\label{app:complete results}

\newcommand{\longHeaderRepresentation}{%
    \textbf{Representation} & 
    \centering\textbf{Graph Query} \newline (single) & 
    \centering\textbf{Graph Query} \newline (multi) & 
    \centering\textbf{Graph Summarization} & 
    \centering\textbf{Knowledge Inference} & 
    \centering\textbf{Path Exploration} & 
    \centering\textbf{Average} \tabularnewline
}

\setlength{\tabcolsep}{2pt} 
\renewcommand{\arraystretch}{1.2}
\tiny 

\begin{longtable}{p{2cm} >{\centering\arraybackslash}p{2cm} >{\centering\arraybackslash}p{2cm} >{\centering\arraybackslash}p{2cm} >{\centering\arraybackslash}p{2cm} >{\centering\arraybackslash}p{2cm} >{\centering\arraybackslash}p{2cm}}

\caption{Response accuracy and token cost per task across different knowledge graph abstractions and flowsheet representations. Tools: ContextRAG is used to query knowledge graphs; image read for flowsheet images; and file read for Proteus File. Comprehensive results for Anthropic and OpenAI models. The response accuracy is the average of LLM-as-judge score: completeness, coherence, correctness, and relatedness.} 
\label{appendix:representation performance} \\

\toprule
\longHeaderRepresentation
\midrule
\endfirsthead

\multicolumn{7}{l}{\footnotesize \ldots App. Table~\ref{appendix:representation performance}, continued from the previous page.} \\
\midrule
\longHeaderRepresentation
\midrule
\endhead

\midrule
\multicolumn{7}{r}{\footnotesize Continued on next page \dots} \\
\endfoot

\bottomrule
\endlastfoot


\multicolumn{7}{c}{\textbf{OpenAI: GPT-5}} \\
\midrule
Graph Abstractions & & & & & & \\
- Conceptual Level & 0.91 / \$0.011 & 1.00 / \$0.033 & 1.00 / \$0.064 & 0.98 / \$0.055 & 0.92 / \$0.027 & 0.94 / \$0.027 \\
- Process Level    & 0.93 / \$0.035 & 0.96 / \$0.060 & 1.00 / \$0.096 & 0.98 / \$0.076 & 0.87 / \$0.063 & 0.93 / \$0.055 \\
- Complete Level   & 0.95 / \$0.086 & 0.75 / \$0.126 & 1.00 / \$0.168 & 0.98 / \$0.126 & 0.82 / \$0.132 & 0.90 / \$0.113 \\
Flowsheet Image    & 0.79 / \$0.002 & 0.61 / \$0.022 & 0.93 / \$0.027 & 0.87 / \$0.020 & 0.67 / \$0.008 & 0.76 / \$0.001 \\
Proteus File       & 0.89 / \$0.152 & 0.74 / \$0.195 & 1.00 / \$0.216 & 0.98 / \$0.191 & 0.87 / \$0.184 & 0.89 / \$0.175 \\
\midrule

\multicolumn{7}{c}{\textbf{OpenAI: GPT-5-mini}} \\
\midrule
Graph Abstractions & & & & & & \\
- Conceptual Level & 0.89 / \$0.002 & 0.95 / \$0.006 & 1.00 / \$0.010 & 0.98 / \$0.008 & 0.88 / \$0.005 & 0.91 / \$0.004 \\
- Process Level    & 0.99 / \$0.007 & 0.98 / \$0.012 & 0.93 / \$0.017 & 0.96 / \$0.013 & 0.79 / \$0.012 & 0.93 / \$0.010 \\
- Complete Level   & 0.94 / \$0.017 & 0.84 / \$0.022 & 0.85 / \$0.023 & 0.99 / \$0.023 & 0.73 / \$0.023 & 0.88 / \$0.021 \\
Flowsheet Image    & 0.79 / \$0.001 & 0.85 / \$0.002 & 0.95 / \$0.005 & 0.91 / \$0.005 & 0.80 / \$0.001 & 0.83 / \$0.002 \\
Proteus File       & 0.90 / \$0.030 & 0.79 / \$0.039 & 1.00 / \$0.040 & 0.99 / \$0.036 & 0.79 / \$0.036 & 0.88 / \$0.034 \\
\midrule

\multicolumn{7}{c}{\textbf{OpenAI: GPT-4o}} \\
\midrule
Graph Abstractions & & & & & & \\
- Conceptual Level & 0.99 / \$0.031 & 0.81 / \$0.024 & 0.60 / \$0.024 & 0.64 / \$0.025 & 0.74 / \$0.025 & 0.83 / \$0.027 \\
- Process Level    & 0.94 / \$0.064 & 0.81 / \$0.071 & 0.60 / \$0.070 & 0.68 / \$0.047 & 0.59 / \$0.067 & 0.77 / \$0.063 \\
- Complete Level   & 0.90 / \$0.165 & 0.70 / \$0.171 & 0.68 / \$0.171 & 0.65 / \$0.115 & 0.57 / \$0.168 & 0.74 / \$0.159 \\
Flowsheet Image    & 0.62 / \$0.001 & 0.39 / \$0.002 & 0.60 / \$0.006 & 0.64 / \$0.005 & 0.56 / \$0.002 & 0.58 / \$0.002 \\
Proteus File       & 0.89 / \$0.300 & 0.60 / \$0.305 & 0.70 / \$0.305 & 0.70 / \$0.204 & 0.55 / \$0.302 & 0.73 / \$0.286 \\
\midrule

\multicolumn{7}{c}{\textbf{OpenAI: GPT-4o-mini}} \\
\midrule
Graph Abstractions & & & & & & \\
- Conceptual Level & 0.98 / \$0.001 & 0.78 / \$0.001 & 0.50 / \$0.001 & 0.63 / \$0.001 & 0.61 / \$0.001 & 0.78 / \$0.001 \\
- Process Level    & 0.91 / \$0.004 & 0.66 / \$0.004 & 0.53 / \$0.004 & 0.61 / \$0.003 & 0.55 / \$0.004 & 0.72 / \$0.004 \\
- Complete Level   & 0.93 / \$0.010 & 0.58 / \$0.010 & 0.50 / \$0.010 & 0.61 / \$0.007 & 0.53 / \$0.010 & 0.71 / \$0.010 \\
Flowsheet Image    & 0.57 / \$0.0001 & 0.33 / \$0.0003 & 0.68 / \$0.0004 & 0.62 / \$0.0003 & 0.55 / \$0.0001 & 0.55 / \$0.0002 \\
Proteus File       & 0.85 / \$0.018 & 0.48 / \$0.018 & 0.58 / \$0.019 & 0.74 / \$0.012 & 0.60 / \$0.018 & 0.71 / \$0.017 \\
\midrule

\multicolumn{7}{c}{\textbf{Anthropic: Claude Opus 4.1}} \\
\midrule
Graph Abstractions & & & & & & \\
- Conceptual Level & 0.90 / \$0.135 & 0.93 / \$0.192 & 0.85 / \$0.203 & 0.78 / \$0.210 & 0.77 / \$0.157 & 0.85 / \$0.162 \\
- Process Level    & 1.00 / \$0.465 & 0.89 / \$0.515 & 0.78 / \$0.526 & 0.83 / \$0.515 & 0.78 / \$0.493 & 0.89 / \$0.489 \\
- Complete Level   & 0.93 / \$1.157 & 0.66 / \$1.207 & 0.80 / \$1.214 & 0.84 / \$1.202 & 0.72 / \$1.175 & 0.83 / \$1.177 \\
Flowsheet Image    & 0.71 / \$0.018 & 0.51 / \$0.061 & 0.65 / \$0.083 & 0.69 / \$0.061 & 0.62 / \$0.038 & 0.66 / \$0.038 \\
Proteus File       & 0.80 / \$1.867 & 0.61 / \$1.768 & 0.85 / \$2.351 & 0.85 / \$2.348 & 0.75 / \$2.320 & 0.78 / \$2.077 \\
\midrule

\multicolumn{7}{c}{\textbf{Anthropic: Claude Sonnet 4}} \\
\midrule
Graph Abstractions & & & & & & \\
- Conceptual Level & 0.96 / \$0.027 & 0.91 / \$0.039 & 0.68 / \$0.041 & 0.81 / \$0.041 & 0.79 / \$0.032 & 0.87 / \$0.033 \\
- Process Level    & 1.00 / \$0.093 & 0.79 / \$0.103 & 0.73 / \$0.105 & 0.82 / \$0.102 & 0.78 / \$0.099 & 0.88 / \$0.098 \\
- Complete Level   & 0.98 / \$0.232 & 0.73 / \$0.243 & 0.73 / \$0.243 & 0.79 / \$0.242 & 0.71 / \$0.238 & 0.84 / \$0.237 \\
Flowsheet Image    & 0.72 / \$0.004 & 0.46 / \$0.013 & 0.60 / \$0.020 & 0.73 / \$0.013 & 0.67 / \$0.009 & 0.68 / \$0.009 \\
Proteus File       & 0.91 / \$0.461 & 0.61 / \$0.473 & 0.78 / \$0.473 & 0.83 / \$0.470 & 0.74 / \$0.464 & 0.81 / \$0.465 \\
\midrule

\multicolumn{7}{c}{\textbf{Anthropic: Claude 3.7 Sonnet}} \\
\midrule
Graph Abstractions & & & & & & \\
- Conceptual Level & 0.95 / \$0.027 & 0.89 / \$0.039 & 0.83 / \$0.040 & 0.79 / \$0.040 & 0.83 / \$0.030 & 0.88 / \$0.032 \\
- Process Level    & 0.98 / \$0.093 & 0.79 / \$0.105 & 0.63 / \$0.105 & 0.78 / \$0.102 & 0.75 / \$0.097 & 0.85 / \$0.097 \\
- Complete Level   & 0.97 / \$0.231 & 0.70 / \$0.241 & 0.63 / \$0.243 & 0.80 / \$0.241 & 0.69 / \$0.236 & 0.82 / \$0.236 \\
Flowsheet Image    & 0.66 / \$0.005 & 0.44 / \$0.011 & 0.63 / \$0.015 & 0.66 / \$0.012 & 0.61 / \$0.007 & 0.62 / \$0.007 \\
Proteus File       & 0.90 / \$0.460 & 0.58 / \$0.473 & 0.80 / \$0.473 & 0.81 / \$0.469 & 0.74 / \$0.464 & 0.81 / \$0.464 \\
\midrule

\multicolumn{7}{c}{\textbf{Anthropic: Claude 3.5 Haiku}} \\
\midrule
Graph Abstractions & & & & & & \\
- Conceptual Level & 0.93 / \$0.007 & 0.81 / \$0.010 & 0.65 / \$0.010 & 0.73 / \$0.009 & 0.72 / \$0.008 & 0.82 / \$0.008 \\
- Process Level    & 0.93 / \$0.024 & 0.85 / \$0.028 & 0.60 / \$0.027 & 0.72 / \$0.027 & 0.66 / \$0.025 & 0.80 / \$0.025 \\
- Complete Level   & 0.95 / \$0.061 & 0.63 / \$0.063 & 0.58 / \$0.064 & 0.66 / \$0.064 & 0.61 / \$0.062 & 0.76 / \$0.062 \\
Flowsheet Image    & 0.58 / \$0.001 & 0.49 / \$0.002 & 0.58 / \$0.003 & 0.68 / \$0.003 & 0.60 / \$0.002 & 0.59 / \$0.002 \\
Proteus File       & 0.85 / \$0.123 & 0.58 / \$0.125 & 0.65 / \$0.126 & 0.73 / \$0.124 & 0.72 / \$0.123 & 0.76 / \$0.123 \\

\end{longtable}
\newcommand{\longHeaderLLM}{%
    \textbf{Tool} & 
    \centering\textbf{Graph Query} \newline (single) & 
    \centering\textbf{Graph Query} \newline (multi) & 
    \centering\textbf{Graph Sum.} & 
    \centering\textbf{Know. Inf.} & 
    \centering\textbf{Path Expl.} & 
    \centering\textbf{Average} \tabularnewline
}

\setlength{\tabcolsep}{3pt} 
\renewcommand{\arraystretch}{1.2}
\tiny 

\begin{longtable}{p{2.5cm} >{\centering\arraybackslash}p{2.0cm} >{\centering\arraybackslash}p{2.0cm} >{\centering\arraybackslash}p{2.0cm} >{\centering\arraybackslash}p{2.0cm} >{\centering\arraybackslash}p{2.0cm} >{\centering\arraybackslash}p{2.0cm}}

\caption{Tool performance across tasks: response accuracy, cost per task (\$), and execution time (s). Comprehensive results for Anthropic, OpenAI, and Ollama models. The response accuracy is the average of LLM-as-judge score: completeness, coherence, correctness, and relatedness.}
\label{appendix:tool_performance_extended} \\

\toprule
\longHeaderLLM
\midrule
\endfirsthead

\multicolumn{7}{l}{\footnotesize \ldots App. Table~\ref{appendix:tool_performance_extended}, continued from the previous page.} \\
\midrule
\longHeaderLLM
\midrule
\endhead

\midrule
\multicolumn{7}{r}{\footnotesize Continued on next page \dots} \\
\endfoot

\bottomrule
\endlastfoot


\multicolumn{7}{c}{\textbf{Anthropic: Claude 3.5 Haiku}} \\
\midrule
ContextRAG & 0.93 / \$0.007 / 6.1s & 0.81 / \$0.010 / 14.3s & 0.65 / \$0.010 / 19.7s & 0.73 / \$0.009 / 16.4s & 0.72 / \$0.008 / 8.1s & 0.82 / \$0.008 / 9.8s \\
VectorRAG & 0.85 / \$0.002 / 5.7s & 0.66 / \$0.003 / 10.8s & 0.53 / \$0.004 / 16.2s & 0.64 / \$0.004 / 16.9s & 0.58 / \$0.003 / 10.7s & 0.71 / \$0.003 / 9.9s \\
PathRAG & 0.84 / \$0.002 / 55.6s & 0.64 / \$0.002 / 139.1s & 0.58 / \$0.004 / 103.3s & 0.67 / \$0.004 / 110.1s & 0.62 / \$0.002 / 108.1s & 0.72 / \$0.002 / 89.3s \\
CypherRAG & 0.51 / \$0.000 / 2.0s & 0.63 / \$0.001 / 4.3s & 0.43 / \$0.001 / 3.5s & 0.54 / \$0.002 / 8.0s & 0.57 / \$0.001 / 3.0s & 0.54 / \$0.001 / 3.5s \\
Multimodal Context & 0.58 / \$0.001 / 7.1s & 0.49 / \$0.002 / 16.9s & 0.58 / \$0.003 / 23.2s & 0.68 / \$0.003 / 21.0s & 0.60 / \$0.002 / 12.2s & 0.59 / \$0.002 / 12.5s \\
Proteus Context & 0.85 / \$0.123 / 88.6s & 0.58 / \$0.125 / 97.9s & 0.65 / \$0.126 / 103.1s & 0.73 / \$0.124 / 97.9s & 0.72 / \$0.123 / 91.3s & 0.76 / \$0.123 / 92.5s \\

\midrule
\multicolumn{7}{c}{\textbf{Anthropic: Claude 3.7 Sonnet}} \\
\midrule
ContextRAG & 0.95 / \$0.027 / 9.2s & 0.89 / \$0.039 / 17.1s & 0.83 / \$0.040 / 24.9s & 0.79 / \$0.040 / 21.3s & 0.83 / \$0.030 / 11.8s & 0.88 / \$0.032 / 13.4s \\
VectorRAG & 0.85 / \$0.007 / 21.2s & 0.59 / \$0.015 / 36.0s & 0.65 / \$0.020 / 58.5s & 0.62 / \$0.019 / 36.2s & 0.61 / \$0.023 / 58.4s & 0.71 / \$0.015 / 36.9s \\
PathRAG & 0.85 / \$0.006 / 59.5s & 0.56 / \$0.009 / 131.2s & 0.60 / \$0.014 / 132.6s & 0.65 / \$0.012 / 112.8s & 0.63 / \$0.009 / 114.2s & 0.72 / \$0.009 / 93.7s \\
CypherRAG & 0.60 / \$0.002 / 4.4s & 0.65 / \$0.006 / 7.7s & 0.63 / \$0.012 / 15.7s & 0.64 / \$0.010 / 13.1s & 0.59 / \$0.003 / 6.0s & 0.61 / \$0.005 / 7.1s \\
Multimodal Context & 0.66 / \$0.005 / 15.0s & 0.44 / \$0.011 / 26.5s & 0.63 / \$0.015 / 34.4s & 0.66 / \$0.012 / 27.5s & 0.61 / \$0.007 / 19.4s & 0.62 / \$0.007 / 20.4s \\
Proteus Context & 0.90 / \$0.460 / 107.9s & 0.58 / \$0.473 / 111.4s & 0.80 / \$0.473 / 110.4s & 0.81 / \$0.469 / 105.4s & 0.74 / \$0.464 / 101.5s & 0.81 / \$0.464 / 106.3s \\

\midrule
\multicolumn{7}{c}{\textbf{Anthropic: Claude Opus 4.1}} \\
\midrule
ContextRAG & 0.90 / \$0.135 / 8.7s & 0.93 / \$0.192 / 29.0s & 0.85 / \$0.203 / 38.7s & 0.78 / \$0.210 / 29.7s & 0.77 / \$0.157 / 17.6s & 0.85 / \$0.162 / 18.1s \\
VectorRAG & 0.83 / \$0.034 / 17.7s & 0.49 / \$0.099 / 54.4s & 0.55 / \$0.120 / 65.2s & 0.67 / \$0.084 / 42.8s & 0.60 / \$0.090 / 46.4s & 0.69 / \$0.068 / 35.6s \\
PathRAG & 0.87 / \$0.031 / 60.0s & 0.61 / \$0.049 / 128.4s & 0.53 / \$0.072 / 119.6s & 0.68 / \$0.065 / 103.5s & 0.59 / \$0.047 / 108.9s & 0.72 / \$0.045 / 90.1s \\
CypherRAG & 0.73 / \$0.013 / 10.0s & 0.73 / \$0.029 / 17.4s & 0.49 / \$0.036 / 35.7s & 0.68 / \$0.047 / 26.4s & 0.62 / \$0.023 / 16.1s & 0.68 / \$0.024 / 16.3s \\
Multimodal Context & 0.71 / \$0.018 / 14.3s & 0.51 / \$0.061 / 34.1s & 0.65 / \$0.083 / 50.1s & 0.69 / \$0.061 / 41.1s & 0.62 / \$0.038 / 25.9s & 0.66 / \$0.038 / 25.6s \\
Proteus Context & 0.80 / \$1.867 / 15.6s & 0.61 / \$1.768 / 24.3s & 0.85 / \$2.351 / 42.3s & 0.85 / \$2.348 / 41.4s & 0.75 / \$2.320 / 30.3s & 0.78 / \$2.077 / 25.8s \\

\midrule
\multicolumn{7}{c}{\textbf{Anthropic: Claude Sonnet 4}} \\
\midrule
ContextRAG & 0.96 / \$0.027 / 7.7s & 0.91 / \$0.039 / 16.0s & 0.68 / \$0.041 / 29.1s & 0.81 / \$0.041 / 22.3s & 0.79 / \$0.032 / 12.9s & 0.87 / \$0.033 / 13.4s \\
VectorRAG & 0.87 / \$0.007 / 8.5s & 0.48 / \$0.020 / 23.9s & 0.55 / \$0.023 / 32.6s & 0.64 / \$0.018 / 26.1s & 0.64 / \$0.019 / 29.0s & 0.71 / \$0.014 / 19.6s \\
PathRAG & 0.85 / \$0.006 / 53.1s & 0.50 / \$0.010 / 119.0s & 0.63 / \$0.014 / 114.4s & 0.68 / \$0.012 / 101.1s & 0.63 / \$0.010 / 102.9s & 0.72 / \$0.009 / 83.9s \\
CypherRAG & 0.73 / \$0.003 / 9.0s & 0.77 / \$0.006 / 15.8s & 0.59 / \$0.006 / 19.5s & 0.65 / \$0.009 / 24.6s & 0.66 / \$0.005 / 15.0s & 0.69 / \$0.005 / 14.3s \\
Multimodal Context & 0.72 / \$0.004 / 11.0s & 0.46 / \$0.013 / 26.5s & 0.60 / \$0.020 / 55.7s & 0.73 / \$0.013 / 36.7s & 0.67 / \$0.009 / 19.9s & 0.68 / \$0.009 / 21.4s \\
Proteus Context & 0.91 / \$0.461 / 11.2s & 0.61 / \$0.473 / 20.2s & 0.78 / \$0.473 / 31.7s & 0.83 / \$0.470 / 24.8s & 0.74 / \$0.464 / 16.8s & 0.81 / \$0.465 / 16.8s \\

\midrule
\multicolumn{7}{c}{\textbf{OpenAI: GPT-5-mini}} \\
\midrule
ContextRAG & 0.89 / \$0.002 / 7.5s & 0.95 / \$0.006 / 34.4s & 1.00 / \$0.010 / 62.5s & 0.98 / \$0.008 / 48.9s & 0.88 / \$0.005 / 24.9s & 0.91 / \$0.004 / 24.3s \\
VectorRAG & 0.90 / \$0.001 / 14.0s & 0.79 / \$0.002 / 20.5s & 0.78 / \$0.004 / 31.9s & 0.87 / \$0.005 / 36.2s & 0.68 / \$0.002 / 34.0s & 0.82 / \$0.002 / 24.4s \\
PathRAG & 0.87 / \$0.001 / 51.6s & 0.65 / \$0.003 / 40.5s & 0.88 / \$0.003 / 24.3s & 0.88 / \$0.005 / 45.9s & 0.79 / \$0.002 / 76.4s & 0.83 / \$0.002 / 54.6s \\
CypherRAG & 0.88 / \$0.000 / 19.4s & 0.81 / \$0.002 / 52.0s & 0.93 / \$0.003 / 52.3s & 0.86 / \$0.005 / 74.9s & 0.84 / \$0.001 / 42.6s & 0.86 / \$0.002 / 39.4s \\
Multimodal Context & 0.79 / \$0.001 / 18.5s & 0.85 / \$0.002 / 42.7s & 0.95 / \$0.005 / 100.4s & 0.91 / \$0.005 / 103.0s & 0.80 / \$0.001 / 44.6s & 0.83 / \$0.002 / 45.6s \\
Proteus Context & 0.90 / \$0.030 / 15.3s & 0.79 / \$0.039 / 96.7s & 1.00 / \$0.040 / 140.5s & 0.99 / \$0.036 / 70.0s & 0.79 / \$0.036 / 64.5s & 0.88 / \$0.034 / 52.1s \\

\midrule
\multicolumn{7}{c}{\textbf{OpenAI: GPT-4o}} \\
\midrule
ContextRAG & 0.99 / \$0.031 / 2.6s & 0.81 / \$0.024 / 13.1s & 0.60 / \$0.024 / 17.4s & 0.64 / \$0.025 / 11.0s & 0.74 / \$0.025 / 5.9s & 0.83 / \$0.027 / 6.7s \\
VectorRAG & 0.80 / \$0.002 / 2.7s & 0.43 / \$0.004 / 3.9s & 0.63 / \$0.004 / 6.0s & 0.54 / \$0.003 / 4.8s & 0.52 / \$0.004 / 4.4s & 0.64 / \$0.003 / 3.8s \\
PathRAG & 0.78 / \$0.003 / 12.5s & 0.53 / \$0.004 / 17.1s & 0.68 / \$0.003 / 4.8s & 0.46 / \$0.003 / 12.7s & 0.51 / \$0.004 / 16.6s & 0.63 / \$0.003 / 13.7s \\
CypherRAG & 0.78 / \$0.001 / 4.7s & 0.64 / \$0.005 / 12.1s & 0.48 / \$0.003 / 11.4s & 0.46 / \$0.002 / 9.4s & 0.42 / \$0.002 / 8.8s & 0.60 / \$0.002 / 7.6s \\
Multimodal Context & 0.62 / \$0.001 / 7.1s & 0.39 / \$0.002 / 12.4s & 0.60 / \$0.006 / 26.7s & 0.64 / \$0.005 / 21.8s & 0.56 / \$0.002 / 11.7s & 0.58 / \$0.002 / 12.2s \\
Proteus Context & 0.89 / \$0.300 / 11.5s & 0.60 / \$0.305 / 35.2s & 0.70 / \$0.305 / 39.2s & 0.70 / \$0.204 / 24.6s & 0.55 / \$0.302 / 20.2s & 0.73 / \$0.286 / 19.8s \\

\midrule
\multicolumn{7}{c}{\textbf{OpenAI: GPT-4o-mini}} \\
\midrule
ContextRAG & 0.98 / \$0.001 / 2.0s & 0.78 / \$0.001 / 10.2s & 0.50 / \$0.001 / 8.7s & 0.63 / \$0.001 / 10.4s & 0.61 / \$0.001 / 5.6s & 0.78 / \$0.001 / 5.5s \\
VectorRAG & 0.82 / \$0.0001 / 3.2s & 0.33 / \$0.0003 / 8.3s & 0.58 / \$0.0001 / 2.8s & 0.61 / \$0.0004 / 12.4s & 0.52 / \$0.0002 / 7.9s & 0.64 / \$0.0002 / 6.4s \\
PathRAG & 0.82 / \$0.0001 / 12.9s & 0.66 / \$0.0003 / 20.2s & 0.50 / \$0.0001 / 2.1s & 0.50 / \$0.0003 / 21.5s & 0.46 / \$0.0003 / 20.4s & 0.64 / \$0.0002 / 16.4s \\
CypherRAG & 0.68 / \$0.0001 / 5.5s & 0.48 / \$0.0002 / 20.3s & 0.50 / \$0.0002 / 13.6s & 0.55 / \$0.0002 / 15.1s & 0.37 / \$0.0001 / 11.0s & 0.54 / \$0.0001 / 10.4s \\
Multimodal Context & 0.57 / \$0.0001 / 7.0s & 0.33 / \$0.0003 / 17.4s & 0.68 / \$0.0004 / 20.4s & 0.62 / \$0.0003 / 26.6s & 0.55 / \$0.0001 / 10.7s & 0.55 / \$0.0002 / 12.9s \\
Proteus Context & 0.85 / \$0.018 / 8.2s & 0.48 / \$0.018 / 24.8s & 0.58 / \$0.019 / 27.3s & 0.74 / \$0.012 / 16.1s & 0.60 / \$0.018 / 13.5s & 0.71 / \$0.017 / 13.6s \\

\midrule
\multicolumn{7}{c}{\textbf{OpenAI: GPT-5}} \\
\midrule
ContextRAG & 0.91 / \$0.011 / 9.6s & 1.00 / \$0.033 / 55.0s & 1.00 / \$0.064 / 124.7s & 0.98 / \$0.055 / 85.8s & 0.92 / \$0.027 / 40.7s & 0.94 / \$0.027 / 40.6s \\
VectorRAG & 0.85 / \$0.008 / 21.6s & 0.73 / \$0.019 / 45.6s & 0.73 / \$0.022 / 33.8s & 0.86 / \$0.033 / 61.3s & 0.68 / \$0.017 / 46.1s & 0.79 / \$0.017 / 37.5s \\
PathRAG & 0.86 / \$0.006 / 62.2s & 0.63 / \$0.010 / 142.1s & 0.83 / \$0.026 / 42.8s & 0.87 / \$0.036 / 112.4s & 0.74 / \$0.017 / 131.2s & 0.80 / \$0.015 / 95.7s \\
CypherRAG & 0.82 / \$0.006 / 48.1s & 0.80 / \$0.010 / 69.8s & 0.95 / \$0.021 / 192.8s & 0.83 / \$0.029 / 199.5s & 0.68 / \$0.010 / 107.3s & 0.79 / \$0.012 / 97.5s \\
Multimodal Context & 0.79 / \$0.002 / 38.2s & 0.61 / \$0.022 / 167.9s & 0.93 / \$0.027 / 231.8s & 0.87 / \$0.020 / 161.3s & 0.67 / \$0.008 / 144.5s & 0.76 / \$0.010 / 109.5s \\
Proteus Context & 0.89 / \$0.152 / 21.2s & 0.74 / \$0.195 / 123.4s & 1.00 / \$0.216 / 212.2s & 0.98 / \$0.191 / 98.9s & 0.87 / \$0.184 / 95.0s & 0.89 / \$0.175 / 73.7s \\

\midrule
\multicolumn{7}{c}{\textbf{Ollama: GPT-OSS 20b}} \\
\midrule
ContextRAG & 0.25 / - / 531.9s & 0.25 / - / 394.6s & 0.20 / - / 757.2s & 0.38 / - / 201.9s & 0.37 / - / 160.7s & 0.30 / - / 379.5s \\
VectorRAG & 0.86 / - / 18.4s & 0.59 / - / 36.0s & 0.55 / - / 50.8s & 0.68 / - / 47.8s & 0.61 / - / 62.4s & 0.72 / - / 38.2s \\
PathRAG & 0.81 / - / 55.2s & 0.58 / - / 91.0s & 0.60 / - / 124.3s & 0.74 / - / 105.0s & 0.58 / - / 94.1s & 0.70 / - / 80.7s \\
CypherRAG & 0.33 / - / 40.2s & 0.40 / - / 88.8s & 0.43 / - / 158.2s & 0.64 / - / 73.1s & 0.56 / - / 102.2s & 0.45 / - / 73.0s \\

\midrule
\multicolumn{7}{c}{\textbf{Ollama: Llama 3.1 8b}} \\
\midrule
ContextRAG & 0.27 / - / 44.2s & 0.23 / - / 55.6s & 0.45 / - / 52.6s & 0.24 / - / 51.0s & 0.33 / - / 54.9s & 0.28 / - / 49.7s \\
VectorRAG & 0.77 / - / 8.5s & 0.34 / - / 17.6s & 0.45 / - / 26.8s & 0.50 / - / 22.0s & 0.51 / - / 11.7s & 0.59 / - / 13.4s \\
PathRAG & 0.79 / - / 65.0s & 0.50 / - / 148.6s & 0.45 / - / 133.2s & 0.52 / - / 124.2s & 0.50 / - / 115.5s & 0.62 / - / 100.0s \\
CypherRAG & - / - / - & - / - / - & - / - / - & - / - / - & - / - / - & - / - / - \\

\midrule
\multicolumn{7}{c}{\textbf{Ollama: Qwen 3 14b}} \\
\midrule
ContextRAG & 0.34 / - / 243.3s & 0.38 / - / 332.8s & 0.73 / - / 329.3s & 0.50 / - / 275.6s & 0.49 / - / 326.8s & 0.43 / - / 284.4s \\
VectorRAG & 0.79 / - / 55.8s & 0.38 / - / 96.7s & 0.40 / - / 89.9s & 0.55 / - / 101.9s & 0.58 / - / 112.9s & 0.63 / - / 84.2s \\
PathRAG & 0.81 / - / 106.8s & 0.48 / - / 158.3s & 0.50 / - / 112.7s & 0.63 / - / 188.7s & 0.60 / - / 158.4s & 0.68 / - / 139.1s \\
CypherRAG & - / - / - & - / - / - & - / - / - & 0.48 / - / 295.1s & 0.55 / - / 257.9s & 0.52 / - / 272.7s \\

\midrule
\multicolumn{7}{c}{\textbf{Ollama: Qwen 3 8b}} \\
\midrule
ContextRAG & 0.27 / - / 182.2s & 0.38 / - / 163.9s & 0.55 / - / 196.8s & 0.45 / - / 147.8s & 0.38 / - / 172.3s & 0.35 / - / 173.0s \\
VectorRAG & 0.83 / - / 35.0s & 0.50 / - / 45.0s & 0.60 / - / 90.6s & 0.59 / - / 71.2s & 0.56 / - / 58.0s & 0.67 / - / 50.7s \\
PathRAG & 0.85 / - / 96.5s & 0.56 / - / 183.5s & 0.58 / - / 232.5s & 0.63 / - / 163.4s & 0.58 / - / 177.2s & 0.70 / - / 144.6s \\
CypherRAG & - / - / - & - / - / - & - / - / - & 0.33 / - / 86.0s & - / - / - & 0.33 / - / 86.0s \\

\midrule
\multicolumn{7}{c}{\textbf{Ollama: Qwen 3 4b}} \\
\midrule
ContextRAG & 0.31 / - / 216.7s & 0.41 / - / 534.6s & 0.53 / - / 319.2s & 0.47 / - / 242.2s & 0.36 / - / 276.7s & 0.37 / - / 275.4s \\
VectorRAG & 0.83 / - / 30.6s & 0.40 / - / 54.3s & 0.50 / - / 115.9s & 0.54 / - / 63.5s & 0.49 / - / 72.0s & 0.63 / - / 53.6s \\
PathRAG & 0.82 / - / 85.9s & 0.54 / - / 178.3s & 0.50 / - / 213.8s & 0.61 / - / 156.6s & 0.59 / - / 189.0s & 0.68 / - / 140.7s \\
CypherRAG & - / - / - & - / - / - & - / - / - & - / - / - & - / - / - & - / - / - \\

\end{longtable}

\end{document}